\documentclass[aps,prd,reprint,superscriptaddress,nofootinbib]{revtex4-1}

\pdfoutput=1
\usepackage[T1]{fontenc}
\usepackage{url}
\usepackage{graphicx}
\usepackage{xspace}
\usepackage[svgnames]{xcolor}
\usepackage{caption}
\usepackage{subcaption}
\usepackage{amsmath}
\usepackage{listings} 
\usepackage[normalem]{ulem}
\usepackage[section]{placeins}
\usepackage{braket}
\usepackage{multirow}
\usepackage{tabulary}

\newcommand{\ccqe}{CCQE\xspace}

\newcommand{\neut}{\textsc{NEUT}\xspace}

\newcommand{\qq}{\ensuremath{Q^{2}}\xspace}
\newcommand{\qqqe}{\ensuremath{Q^{2}_{\textrm{QE}}}\xspace}
\newcommand{\pf}{\ensuremath{p_{\mathrm{F}}}\xspace}

\newcommand{\ndof}{\ensuremath{N_{\mathrm{DOF}}}\xspace}

\newcommand{\Enu}{\ensuremath{E_{\nu}}\xspace}
\newcommand{\ma}{\ensuremath{M_{\textrm{A}}}\xspace}

\newcommand{\mnn}{multinucleon--neutrino\xspace}

\newcommand{\mb}{MiniBooNE\xspace}
\newcommand{\minerva}{MINER\ensuremath{\nu}A\xspace}

\def\bracketbar{\hbox{\kern-9pt\raise1pt%
         \hbox{{\tiny(}{\lower1.5pt\hbox{\bf--}}{\tiny)}}}}

\begin{document}

\title{Testing CCQE and 2p2h models in the \neut neutrino interaction generator with published datasets from the \mb and \minerva experiments}
\date{\today}
\author{C.~Wilkinson}
\email[Corresponding author, ]{callum.wilkinson@lhep.unibe.ch}
\affiliation{University of Bern, Albert Einstein Center for Fundamental Physics, Laboratory for High Energy Physics (LHEP), Bern, Switzerland}
\affiliation{University of Sheffield, Department of Physics and Astronomy, Sheffield, United Kingdom}
\author{R.~Terri}
\email[Corresponding author, ]{r.terri@qmul.ac.uk}
\affiliation{Queen Mary University of London, School of Physics and Astronomy, London, United Kingdom}
\author{C.~Andreopoulos}
\affiliation{STFC, Rutherford Appleton Laboratory, Harwell Oxford,  and  Daresbury Laboratory, Warrington, United Kingdom}
\affiliation{University of Liverpool, Department of Physics, Liverpool, United Kingdom}
\author{A.~Bercellie}
\affiliation{University of Rochester, Department of Physics and Astronomy, Rochester, New York, USA}
\author{C.~Bronner}
\affiliation{Kavli Institute for the Physics and Mathematics of the Universe (WPI), The University of Tokyo Institutes for Advanced Study, University of Tokyo, Kashiwa, Chiba, Japan}
\author{S.~Cartwright}
\affiliation{University of Sheffield, Department of Physics and Astronomy, Sheffield, United Kingdom}
\author{P.~de~Perio}
\altaffiliation[Present address: ]{Columbia University, Physics Department, New York, New York 10027, USA}
\affiliation{University of Toronto, Department of Physics, Toronto, Ontario, Canada}
\author{J.~Dobson}
\altaffiliation[Present address: ]{University College London, Department of Physics and Astronomy, London, United Kingdom}
\affiliation{Imperial College London, Department of Physics, London, United Kingdom}
\author{K.~Duffy}
\affiliation{Oxford University, Department of Physics, Oxford, United Kingdom}
\author{A.P.~Furmanski}
\altaffiliation[Present address: ]{University of Manchester, School of Physics and Astronomy, Manchester, United Kingdom}
\affiliation{University of Warwick, Department of Physics, Coventry, United Kingdom}
\author{L.~Haegel}
\affiliation{University of Geneva, Section de Physique, DPNC, Geneva, Switzerland}
\author{Y.~Hayato}
\affiliation{Kavli Institute for the Physics and Mathematics of the Universe (WPI),Todai Institutes for Advanced Study, University of Tokyo, Kashiwa, Chiba, Japan}
\affiliation{University of Tokyo, Institute for Cosmic Ray Research, Kamioka Observatory, Kamioka, Japan}
\author{A.~Kaboth}
\affiliation{Royal Holloway University of London, Department of Physics, Surrey, United Kingdom}
\affiliation{STFC, Rutherford Appleton Laboratory, Harwell Oxford,  and  Daresbury Laboratory, Warrington, United Kingdom}
\author{K.~Mahn}
\affiliation{Michigan State University, Department of Physics and Astronomy, East Lansing, Michigan, U.S.A.}
\author{K.S.~McFarland}
\affiliation{University of Rochester, Department of Physics and Astronomy, Rochester, New York, USA}
\author{J.~Nowak}
\affiliation{Lancaster University, Physics Department, Lancaster, United Kingdom}
\author{A.~Redij}
\affiliation{University of Bern, Albert Einstein Center for Fundamental Physics, Laboratory for High Energy Physics (LHEP), Bern, Switzerland}
\author{P.~Rodrigues}
\affiliation{University of Rochester, Department of Physics and Astronomy, Rochester, New York, USA}
\author{F.\,S\'anchez}
\affiliation{Institut de Fisica d'Altes Energies (IFAE), The Barcelona Institute of Science and Technology, Campus UAB, Bellaterra (Barcelona) Spain}
\author{J.D.~Schwehr}
\affiliation{Colorado State University, Department of Physics, Fort Collins, Colorado, U.S.A.}
\author{P.~Sinclair}
\affiliation{Imperial College London, Department of Physics, London, United Kingdom}
\author{J.T.~Sobczyk}
\affiliation{Wroc\l{}aw University, Faculty of Physics and Astronomy, Wroc\l{}aw, Poland}
\author{P.~Stamoulis}
\affiliation{IFIC (CSIC \& University of Valencia), Valencia, Spain}
\author{P.~Stowell}
\affiliation{University of Sheffield, Department of Physics and Astronomy, Sheffield, United Kingdom}
\author{R.~Tacik}
\affiliation{University of Regina, Department of Physics, Regina, Saskatchewan, Canada}
\affiliation{TRIUMF, Vancouver, British Columbia, Canada}
\author{L.~Thompson}
\affiliation{University of Sheffield, Department of Physics and Astronomy, Sheffield, United Kingdom}
\author{S.~Tobayama}
\affiliation{University of British Columbia, Department of Physics and Astronomy, Vancouver, British Columbia, Canada}
\author{M.O.~Wascko}
\affiliation{Imperial College London, Department of Physics, London, United Kingdom}
\author{J.~\.Zmuda}
\affiliation{Wroc\l{}aw University, Faculty of Physics and Astronomy, Wroc\l{}aw, Poland}

\begin{abstract}
The \mb large axial mass anomaly has prompted a great deal of theoretical work on sophisticated Charged Current Quasi-Elastic (CCQE) neutrino interaction models in recent years. As the dominant interaction mode at T2K energies, and the signal process in oscillation analyses, it is important for the T2K experiment to include realistic CCQE cross section uncertainties in T2K analyses. To this end, T2K's Neutrino Interaction Working Group has implemented a number of recent models in \neut, T2K's primary neutrino interaction event generator. In this paper, we give an overview of the models implemented, and present fits to published $\nu_\mu$ and $\overline{\nu}_\mu$ CCQE cross section measurements from the \mb and \minerva experiments. The results of the fits are used to select a default cross section model for future T2K analyses, and to constrain the cross section uncertainties of the model. We find a model consisting of a modified relativistic Fermi gas model and multinucleon interactions most consistently describes the available data.
\end{abstract}
\maketitle

\section{Introduction}\label{sec:intro}

Charged Current Quasi-Elastic (CCQE) scattering ($\nu_{\mu} + n \rightarrow p + \mu^{-}$) is the dominant neutrino interaction process for muon (anti)neutrinos impinging on a nuclear target at neutrino energies on the order of 1\,GeV. Because CCQE is a two-body process and the incoming neutrino direction is known for an accelerator experiment, a reasonable approximation of the neutrino energy can be calculated using only the outgoing lepton kinematics. Because of this, CCQE is the preferred signal process for neutrino oscillation experiments which generally require some handle on the incoming neutrino energy to extract neutrino oscillation parameters due to $\nu_\mu^{\bracketbar}$ disappearance or $\nu_e^{\bracketbar}$ appearance in this energy region. However, nuclear effects and interactions which are not distinguishable from CCQE in the final state bias or smear the reconstructed neutrino energy, so a good understanding of these effects is important.

Neutrino interaction generators typically use the relativistic Fermi gas (RFG) model of the nucleus for all neutrino-nucleus interactions because of its simplicity. In the RFG model, all possible nucleon momentum states are filled up to the Fermi momentum, there is a constant binding energy required to separate the nucleon from the nucleus, and the neutrino interacts with a single bound nucleon. Neutrino-nucleon CCQE scattering for free nucleons is described by the Llewellyn-Smith formalism~\cite{llewellyn-smith}, which has been extended to cover neutrino-nucleus CCQE scattering in the Smith-Moniz RFG model~\cite{smith-moniz}, where nucleons bound within the nucleus are described by the RFG nuclear model. The only parameter of the weak current or in the RFG model which is not well constrained by electron scattering data~\cite{review, moniz:fg_escat} is the axial mass, \ma. Results from a global analysis of neutrino-deuterium scattering experiments and pion electroproduction data find $\ma = 1.00\pm0.02$\,GeV/$c^{2}$~\cite{bba03}, which is consistent with other analyses~\cite{axialStructure, kuzmin_2006, kuzmin_2008}. These results are also consistent with high energy neutrino beam experiments on heavy nuclear targets~\cite{NOMAD_CCQE}.\\
\indent Recent differential CCQE cross section results from the \mb collaboration~\cite{mb-ccqe-2010, mb-ccqe-antinu-2013} are significantly higher than expectation, which can only be accounted for in the framework of the Smith-Moniz RFG model by inflating the axial mass, giving rise to the term ``\mb large axial mass anomaly''. This came after an earlier large axial mass measurement by K2K~\cite{k2kCCQE}, which reported a value of $\ma=1.20\pm0.12$\,GeV/$c^{2}$. Both experiments exhibited not only a larger-than-expected axial mass, but also a supression of low-\qq events relative to the expection from the Smith-Moniz RFG model. Other experiments using heavy nuclear targets with beam energies in the few-GeV region have also measured cross sections which are consistent with an inflation of the axial mass~\cite{t2kCCQE, ingridCCQE, minosCCQE, Wascko:2008zz}, although these results do not paint a coherent picture. More recently, the \minerva experiment~\cite{minerva-nu-ccqe,minerva-antinu-ccqe}, which is at a somewhat higher neutrino energy than \mb, has shown good agreement with the Smith-Moniz RFG model with \ma = 1.00\,GeV/$c^{2}$, but requires an enhancement to the transverse component of the cross section, an effect also seen in electron-nucleus scattering~\cite{tem}. These inconsistent results present a considerable challenge to neutrino oscillation experiments which need to be able to model their signal processes well.

Recent theoretical efforts which have attempted to resolve the ``MiniBooNE large axial mass anomaly'' have focussed on two main areas: more sophisticated descriptions of the initial ground state of the nucleus; and additional nuclear effects, such as multinucleon interaction models, which go beyond interactions with a single nucleon within the nucleus. The combination of these models would allow for a consistent picture of an axial mass close to 1.00\,GeV/$c^{2}$, with a suppressed cross section at low-\qq and larger cross section at higher-\qq relative to a simple RFG model. Comprehensive reviews of available CCQE cross section models can be found in References~\cite{review, Morfin:2012kn, garvey_review_2014, hayato_review_2014}.

More sophisticated descriptions of the initial state of the nucleus than the RFG model provides are available from a number of authors~\cite{sf, ankowski_SF, butkevich_2009, eff-sf}. These models, generically referred to as Spectral Functions (SF), have a more realistic nucleon momentum distribution taking into account the shell structure of the nucleus and correlated pairs of nucleons within the nucleus, and have non-constant binding energies. Note that although these models include correlations between nucleons in the initial state, they still use the impulse approximation and only consider interactions with a single nucleon. More complex models which go beyond the simple picture of non-interacting fermions are available~\cite{leitner_2009, madrid_2003, meucci_2003, lovato_2013, pandey_2014}. However, with the exception of the GiBUU interaction model~\cite{gibuu, leitner_2009}, these are not currently implemented in neutrino interaction generators. In these models, a mean field potential due to the presence of other nucleons within the nucleus is calculated, which will generally depend on the position and momentum of the struck nucleon. These models are not discussed further as they cannot be easily implemented in the \neut neutrino interaction generator~\cite{hayato:neut}.

Although alternative nuclear models modify the cross section as a function of the outgoing lepton kinematics significantly, they do not change the total CCQE cross section significantly as a function of neutrino energy~\cite{garvey_review_2014}. Additional nuclear effects are also likely to be required to explain the current global dataset. Multinucleon interaction (2p2h) models such as those from Nieves {\it et al.}~\cite{nieves, nievesExtension} and Martini {\it et al.}~\cite{martini} go beyond the impulse approximation and include diagrams where two nucleons are involved in the interaction. This adds significant strength to the CCQE-like cross section and explains the difference in normalization observed in the \mb data, which was previously modelled with a large axial mass~\cite{nieves_MB_nu_2011, martini_MB_nu_2011, nieves_MB_anu_2013, martini_MB_anu_2013}. Because these 2p2h models are not two-body processes, they are expected to lead to significant biases in the neutrino energy reconstruction from the outgoing lepton which assumes CCQE kinematics~\cite{coloma_erec, ankowski_erec}. Additionally, the Random Phase Approximation (RPA) is a nuclear screening effect that modifies the propagator for interactions in nuclear matter~\cite{Jachowicz:1998fn, Jachowicz:2002rr, nieves, martini} and has a significant effect on the differential cross section as a function of \qq, suppressing the cross section in the low-\qq region and enhancing the cross section for $\qq\gtrsim 0.5$\,GeV$^{2}$. RPA needs to be included, both in interactions with a single nucleon (1p1h) and those from 2p2h calculations, to find good agreement with data. Note that both Nieves and Martini calculations are performed in the context of a local Fermi gas (LFG) model, where the Fermi momentum depends on the local nuclear density, so improvements to the initial state models of the nucleus and improvements to the CCQE interaction models cannot necessarily be combined easily.\\
\indent Whilst there have been rapid experimental and theoretical developments relating to CCQE cross sections, new nuclear models and nuclear effects have only recently been implemented into neutrino interaction generators or confronted with neutrino-nucleus scattering data, and no consistent picture has yet emerged. It is not clear which models fit the global data best, and where the deficiencies now lie, which should be a serious concern for neutrino oscillation experiments. This paper shows the effect of fitting current CCQE and multinucleon models to the \mb~\cite{mb-ccqe-2010, mb-ccqe-antinu-2013} and \minerva~\cite{minerva-nu-ccqe, minerva-antinu-ccqe} datasets to a variety of models implemented in \neut by members of T2K's Neutrino Interactions Working Group (NIWG). Previous constraints on the CCQE model produced by the NIWG and used in T2K oscillation analyses only considered an RFG model, and recommended the \neut default central value for the axial mass \ma = 1.21 GeV$/c^{2}$ based on the value found by the K2K experiment~\cite{k2kCCQE}, with an error large enough to cover fits to the \mb neutrino mode CCQE dataset~\cite{mb-ccqe-2010}, as is fully described in Reference~\cite{Abe:2015awa}. This work improves on the previous situation by including more sophisticated effects proposed to explain the large axial mass anomaly, and by using all of the newly available CCQE data to constrain all model parameters without reference to the default \neut model.

The models which have been implemented in the \neut generator are discussed in Section~\ref{sec:neut} and Section~\ref{sec:nuwro} discusses cross-generator validation. Section~\ref{sec:extdata} gives a brief overview of the \mb and \minerva data used in the fit.  The nominal \neut predictions for these datasets are shown in Section~\ref{sec:mcpred} for a variety of models.  Section~\ref{sec:fitprod} discusses the fit procedure. The results of fake data studies and the fit to external data are given in Section~\ref{sec:results}. In Section~\ref{sec:discussion} we interpret the results and discuss possible implications in cross section and neutrino oscillation analyses and Section~\ref{sec:summary} summarizes the results.

\section{Interaction models in \neut}
\label{sec:neut}

The motivation for, and an overview of, new CCQE models has already been discussed. This section will briefly outline the important technical details of the models as implemented in \neut, and highlight any caveats that should be borne in mind when fitting with them.  The models used in the fits include the SF model, \mnn interactions, and RPA.

\begin{figure}[htb]
  \centering
  \includegraphics[width=0.9\columnwidth]{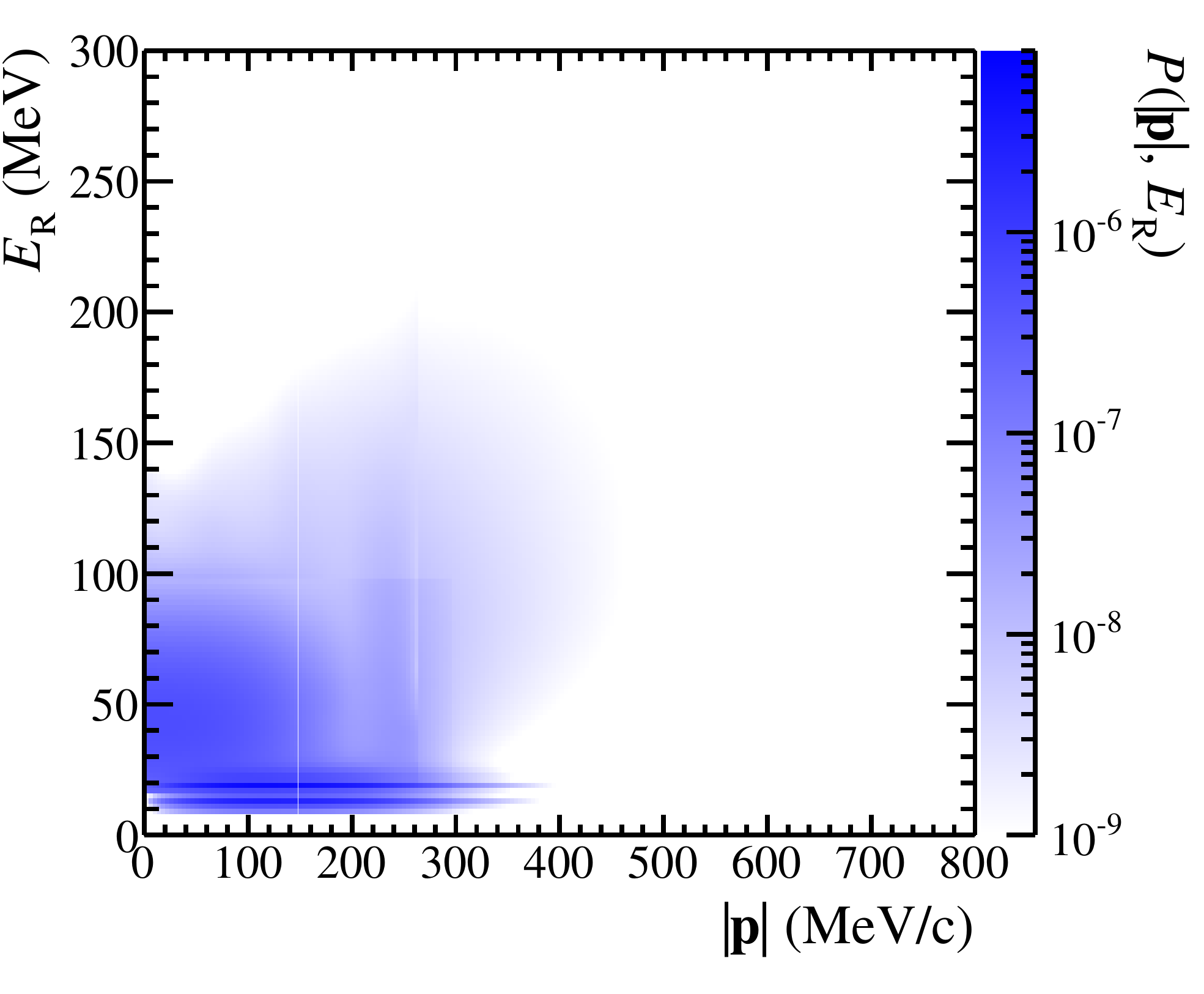}
  \caption{The probability distribution for initial state protons within an oxygen nucleus for Benhar's SF model~\cite{sf} as a function of the removal energy ($E_{\mathrm{R}}$) and the magnitude of the nucleon momentum ($|\mathbf{p}|$). The SF is normalized such that the integral of this distribution is 1.}\label{fig:SF_2D_distribution}
\end{figure}

\begin{figure}[htb]
  \centering
  \includegraphics[trim=0cm 1.2cm 0cm 2.8cm, clip=true, width=0.9\columnwidth]{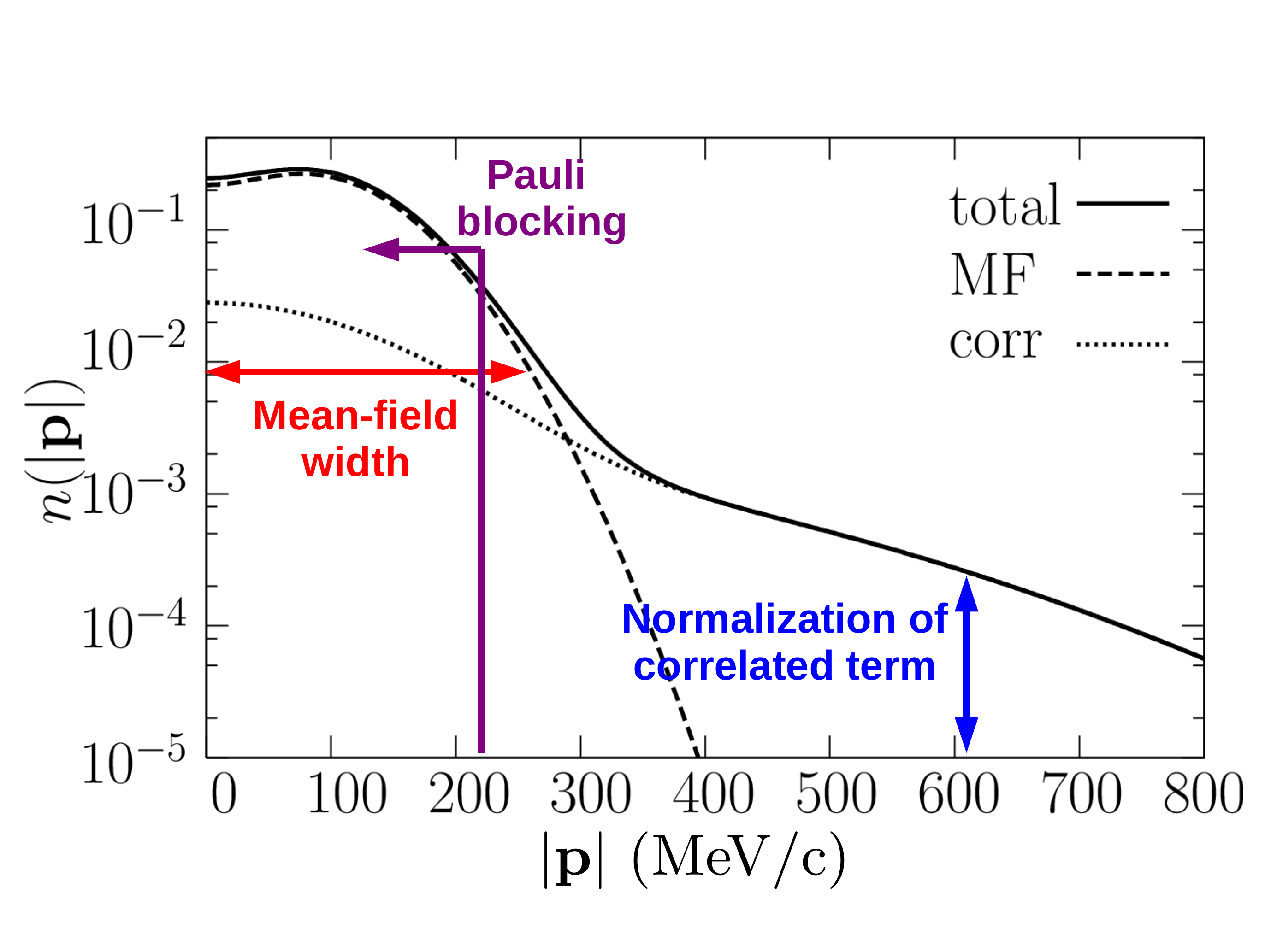}
  \caption{SF parameters in \neut that may be modified on the SF initial state momentum distribution. This figure has been adapted from Reference~\cite{ciofi_1996}.}\label{fig:SF_dials}
\end{figure}
The \neut implementation of the SF model from Omar Benhar and collaborators~\cite{sf} is described fully in Reference~\cite{andy_thesis}. Although SF is a generic term, in this work it will specifically refer to the Benhar SF. The model information is all encoded in the initial state nucleon distribution shown in Figure~\ref{fig:SF_2D_distribution}. Pauli blocking is implemented as a hard cut-off: final state nucleons with momenta less than the Fermi momentum $\pf^{\mbox{\scriptsize{SF}}}$ are forbidden. There are two terms in the SF model: a short range correlation term, which extends to higher initial state nucleon momenta, and a mean field term, which contributes the main peak at lower momenta. These terms can be seen in Figure~\ref{fig:SF_dials}, where the two-dimensional SF in terms of the removal energy and initial state nucleon momentum has been projected on to the momentum axis. There are three parameters in \neut which modify the SF as illustrated in Figure~\ref{fig:SF_dials}. The default values for these parameters are given in Table~\ref{tab:SF_params}. The mean field width and normalization of the correlation term are well-constrained by electron--scattering data~\cite{andy_thesis} and have little effect on the shape or normalization of the cross section. Thus, they are not considered further in this work. Pauli blocking is modified by changing the Fermi momentum in the fits. It should be noted that in the RFG model, the Fermi momentum defines the Pauli blocking, but {\it also} modifies the width of the initial state nucleon distribution. As a result, changing $\pf^{\mbox{\scriptsize{RFG}}}$ affects a wide range of \qq, whereas changing $\pf^{\mbox{\scriptsize{SF}}}$ only affects very low \qq events.

The \mnn (2p2h) model from Nieves {\it et al.}~\cite{nieves, nievesExtension} has been implemented in \neut as described in Reference~\cite{peterthesis}. The cross section as a function of neutrino energy and the outgoing lepton kinematics was made available by the authors of the model and is implemented as a series of lookup tables for various nuclear targets and neutrino species. The tables provided had hadronic variables integrated out, so a generic model based on Reference~\cite{multinucleon} for simulating the initial and final hadronic states was used for generating \neut events\footnote{This model simply enforces energy and momentum conservation, treats initial nucleons as uncorrelated and drawn from a local Fermi gas model, and shares momentum equally between final state nucleons~\cite{multinucleon}.}. The discrepancy between the leptonic and hadronic simulation makes the current \neut implementation of the Nieves model inadequate for comparisons with experimental measurements of the final state hadrons from \ccqe events (such as can be found in Reference~\cite{minerva_ccqe_p_2014}). For this reason, only leptonic measurements are used in this analysis. As the Nieves model is very complex, the current \neut implementation does not allow fundamental model parameters to be changed. For simplicity, only a simple scaling parameter which changes the normalization of 2p2h events has been considered in this analysis. Note that the Nieves 2p2h model included $\pi$-less $\Delta$ decay contributions, where a nucleon excited into a $\Delta(1232)$ resonance decays without producing a pion~\cite{oset-fsi, singh_1998}. Contributions from $\pi$-less $\Delta$ decay were previously implemented in \neut and other generators, and have been treated as an intrinsic background in CCQE selections and corrected for. This leads to complications when comparing the full Nieves model to CCQE cross section measurements.

\begin{figure}[htb]
  \centering
  \begin{subfigure}[t]{0.9\columnwidth}
    \includegraphics[width=\textwidth]{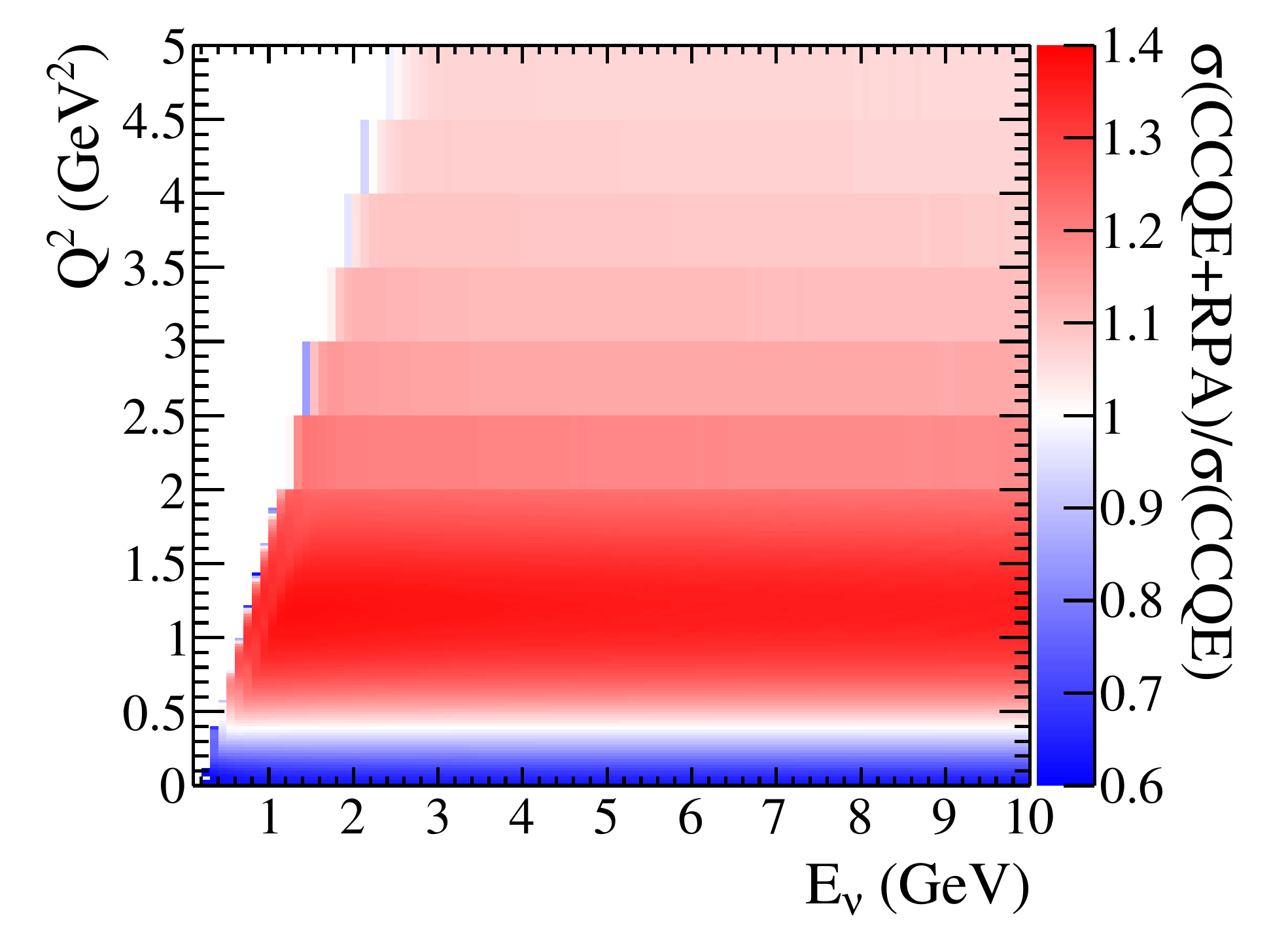}
    \caption{$\nu_{\mu}$ -- $^{12}$C}
  \end{subfigure}
  \begin{subfigure}[t]{0.9\columnwidth}
    \includegraphics[width=\textwidth]{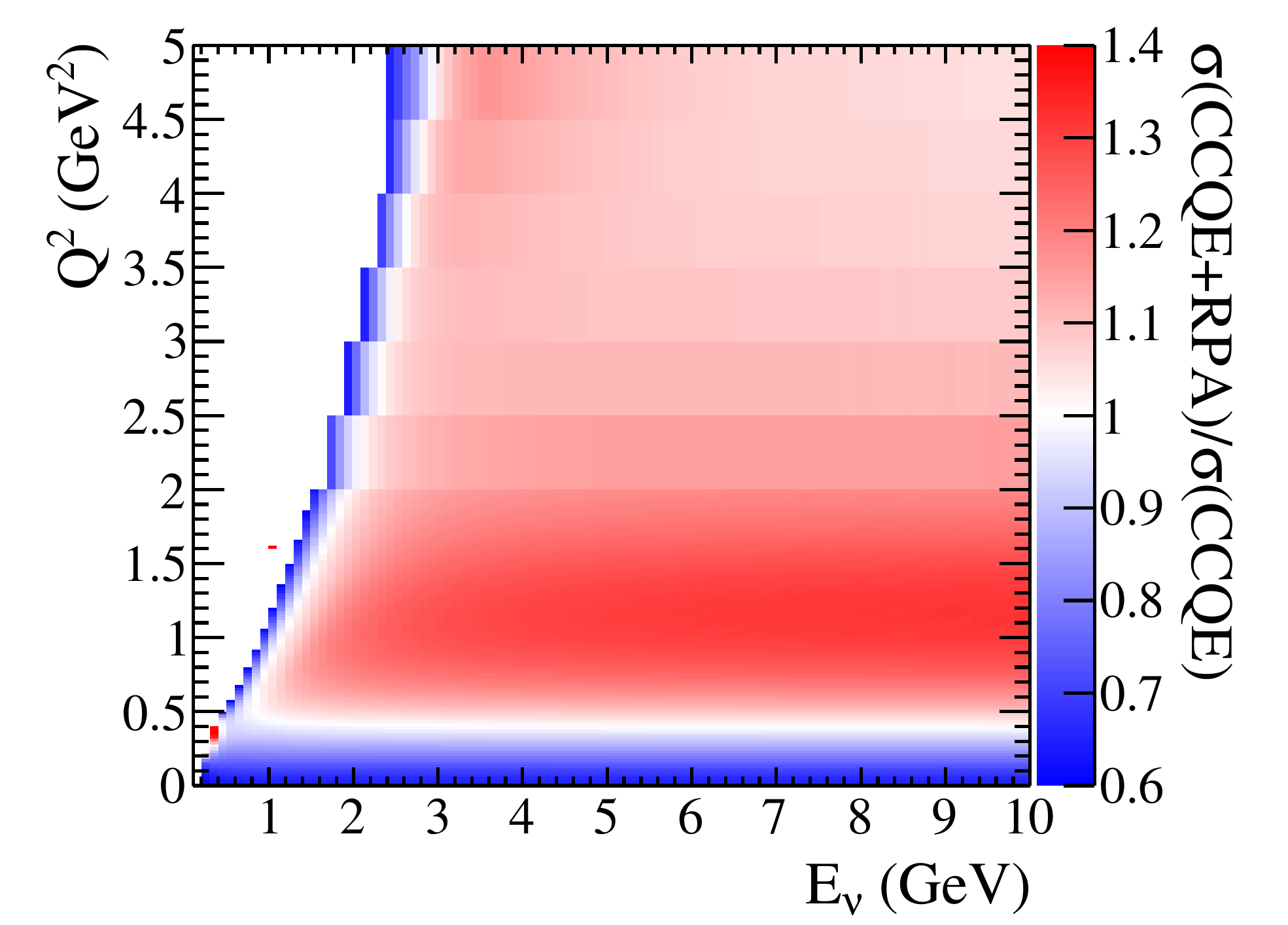}
    \caption{$\bar{\nu}_{\mu}$ -- $^{12}$C}
  \end{subfigure}
  \caption{The ratio of the \ccqe cross section including the non-relativistic RPA model to the \ccqe cross section without RPA, shown for both muon neutrino and muon antineutrino interactions on carbon. An enhancement of the ratio can be seen at high \qq, and a suppression can be seen at low \qq (and close to the kinematic boundary for antineutrinos). These \Enu and \qq dependent tables are used in \neut to apply the RPA model. For these plots, an axial mass value of \ma = 1.01 GeV$/c^{2}$ was used.}\label{fig:RPA_ratio_NEUT}
\end{figure}
\begin{figure}[htb]
  \centering
  \begin{subfigure}[t]{0.9\columnwidth}
    \includegraphics[width=\textwidth]{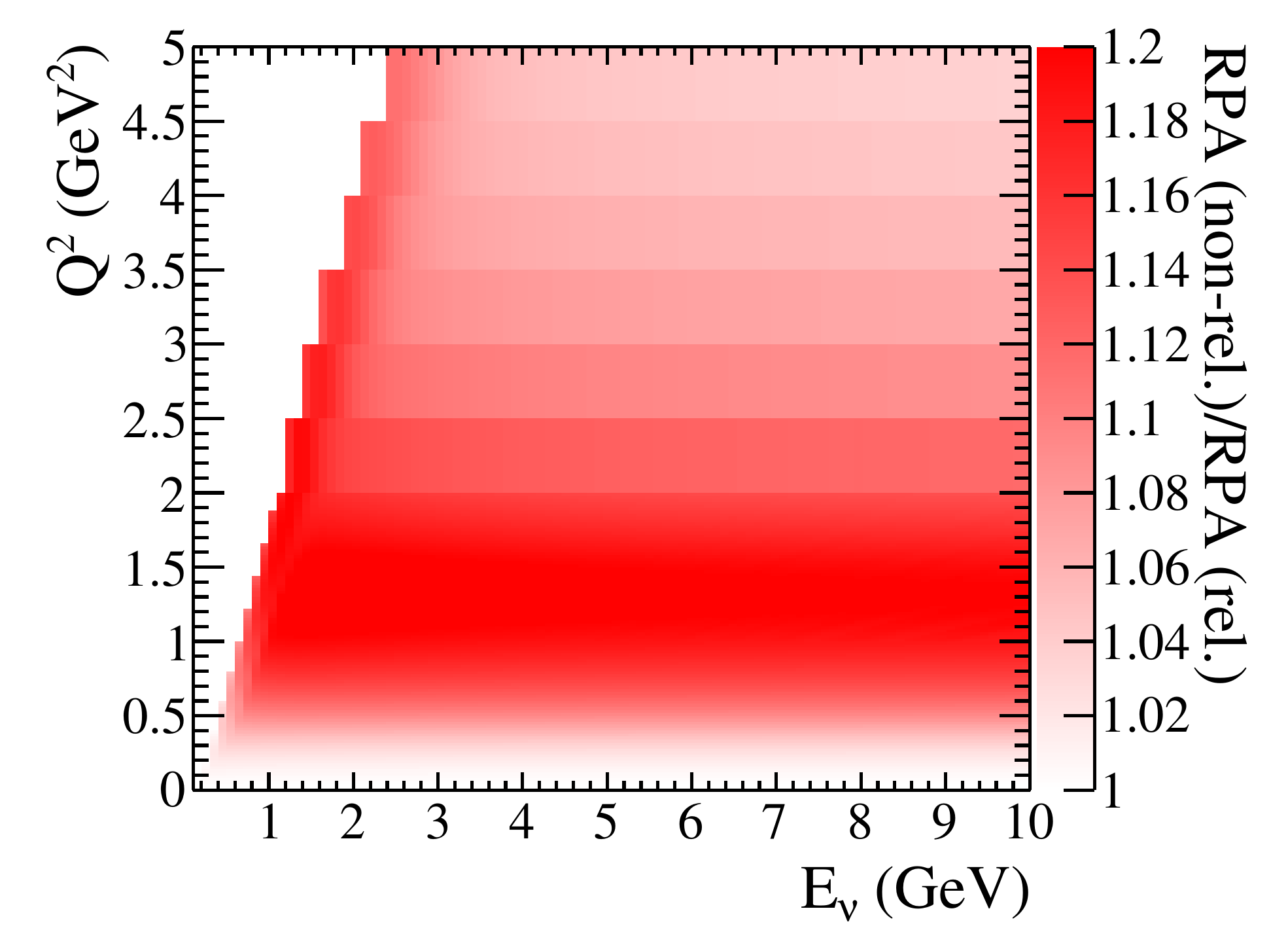}
    \caption{$\nu_{\mu}$ -- $^{12}$C}
  \end{subfigure}
  \begin{subfigure}[t]{0.9\columnwidth}
    \includegraphics[width=\textwidth]{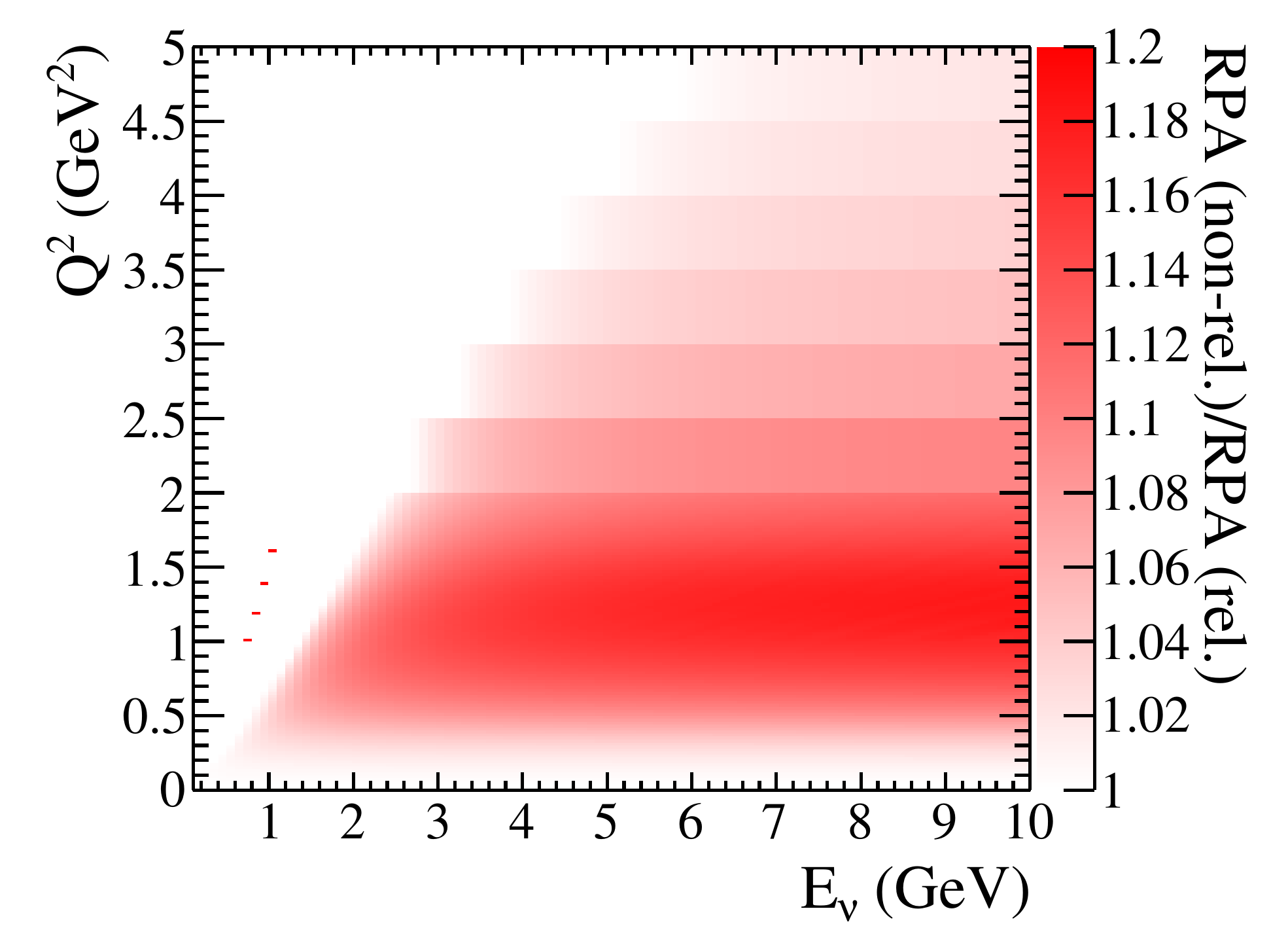}
    \caption{$\bar{\nu}_{\mu}$ -- $^{12}$C}
  \end{subfigure}
  \caption{The ratio of the non-relativistic RPA correction to the relativistic RPA correction, shown for both muon neutrino and muon antineutrino interactions on carbon. These \Enu and \qq dependent tables are used to reweight \neut events from one RPA model to the other. By default, \neut events are generated with the non-relativistic RPA model.}\label{fig:RPA_types_ratio_NEUT}
\end{figure}

RPA~\cite{nieves} is implemented into \neut as a modification to the 1p1h cross section as a function of \Enu and \qq. Figure~\ref{fig:RPA_ratio_NEUT} shows the ratio of the Nieves 1p1h cross section with RPA included over the bare 1p1h cross section; these two-dimensional tables of the ratio were supplied by the authors of Reference~\cite{nieves} and are used to apply the RPA correction in \neut. The Nieves RPA calculation uses the {\it local} Fermi gas nuclear model, and \neut only has a {\it global} Fermi gas model implemented for 1p1h interactions, but the authors of the RPA calculation have noted~\cite{nievesExtension} that the same ratio can be applied, with reasonable precision, to a global Fermi gas. Because of the model dependence, the same ratios cannot be applied to modify the 1p1h interactions calculated with a SF model, and no RPA calculation performed in the context of SF nuclear model is available. Two different RPA calculations are available from the same authors, termed relativistic and non-relativistic, which affect the quenching of the RPA at high \qq ($\gtrsim 0.5$ GeV$^{2}$). The ratio of non-relativistic to relativistic RPA is shown in Figure~\ref{fig:RPA_types_ratio_NEUT}. Both RPA models are investigated in this analysis as there is no guidance on which model is more physical. The `stray' points in Figures~\ref{fig:RPA_ratio_NEUT} and~\ref{fig:RPA_types_ratio_NEUT} are artifacts from the authors of the RPA model, who provided the data used to produce these figures. The cause of these artifacts is unknown, but as these points lie outside the kinematically allowed region of (\Enu, \qq) space, they do not affect the RPA implementation in \neut as no event outside this region can be generated.

With these different ingredients, three distinct candidate CCQE models are available in \neut, which are all considered in this work:
\begin{enumerate}
        \item RFG+relativistic RPA+2p2h
        \item RFG+non-relativistic RPA+2p2h
        \item SF+2p2h.
\end{enumerate}
The default values for all variable model parameters are listed in Table~\ref{tab:RFG_params} and Table~\ref{tab:SF_params} for both RFG+RPA+2p2h models and SF+2p2h, respectively.

It should be noted that there are deficiencies for both models as currently implemented in \neut. The RFG+RPA+2p2h model is very like the full Nieves model as both the RPA and 2p2h calculations are taken from it. However, the Nieves model consistently uses a {\it local} Fermi gas, whereas \neut uses a {\it global} Fermi gas model for the 1p1h calculation. Currently there is no ability to vary the value of \ma used in the Nieves model prediction as implemented in \neut, making the fits slightly inconsistent in this regard\footnote{The value of the axial mass used for the 2p2h contribution to the cross section was fixed to $\ma = 1.01$ GeV$/c^{2}$.}. The SF+2p2h model has no RPA correction applied, which is physically inconsistent as the 2p2h enhancement is used (both corrections are due to complications in heavy nuclear targets). As previously noted, no RPA calculation appropriate for a SF model is currently available, so this inconsistency is unavoidable. The nuclear models used for the 1p1h calculation (SF) and the 2p2h calculation (LFG) are also inconsistent, and it has been remarked that the short range correlations included the SF nuclear model may be the same as some contributions to the Nieves 2p2h interaction model, so some contributions may be included twice.

Additionally, the Effective Spectral Function (ESF)~\cite{eff-sf, eff-sf_proc} has been implemented in \neut as described in Reference~\cite{my_thesis}, and is included for comparison with the other nominal models in Section~\ref{sec:mcpred}. The ESF enforces agreement with the longitudinal response function extracted from electron scattering data by modifying the initial state nucleon momentum distribution (using a simple parametrization of the Benhar SF model), and should be used with the Transverse Enhancement Model (TEM), which parametrizes the observed discrepancy between the longitudinal and transverse response functions extracted from electron scattering data as an enhancement to the magnetic form factor~\cite{tem}. By construction, the ESF+TEM agrees with elastic electron scattering data, and is extended to neutrino scattering data by modifying the Llewellyn-Smith interaction formalism for nucleons bound in a nucleus described by the ESF (and with the modified magnetic form factor from the TEM). This model was implemented too late to be a candidate model for the T2K oscillation analysis, and is not considered further in the fitting work described in this paper.

\begin{table}[htb]
\centering
{\renewcommand{\arraystretch}{1.2}
\begin{tabular}{c|c}
        \hline\hline
        Model parameter & ~~~~~\neut default value~~~~~ \\
        \hline
        \ma & 1.01\,GeV/$c^{2}$ \\
        Fermi momentum, $\pf^{\mathrm{SF}}$ & 209\,MeV/$c$ \\
        Mean-field width & 200\,MeV/$c$ (Benhar nominal~\cite{sf}) \\
        Norm. of the & Benhar nominal~\cite{sf} \\
        correlation term & (correlated tail $\sim$20\% of total) \\
        2p2h normalization & 100\% Nieves model~\cite{nieves, nievesExtension} \\
        Axial form factor & Dipole \\
        Vector form factors & BBBA05~\cite{bbba05} \\
        \hline\hline
\end{tabular}}
\caption{Nominal model parameters for the SF+2p2h model.}\label{tab:SF_params}
\end{table}
\begin{table}[htb]
\centering
{\renewcommand{\arraystretch}{1.2}
\begin{tabular}{c|c}
        \hline\hline    
        Model parameter & ~~~~~\neut default value~~~~~ \\ 
        \hline
        \ma & 1.01\,GeV/$c^{2}$ \\
        Fermi momentum, $\pf^{\mathrm{RFG}}$ & 217\,MeV/$c$ \\
        RPA & Nieves relativistic or \\
         & non-relativistic model~\cite{nieves} \\
        2p2h normalization & 100\% Nieves model~\cite{nieves, nievesExtension} \\
        Axial form factor & Dipole \\
        Vector form factors & BBBA05~\cite{bbba05}\\
        \hline\hline
\end{tabular}}
\caption{Nominal model parameters for the relativistic and non-relativistic RFG+RPA+2p2h models.}\label{tab:RFG_params}
\end{table}
We note that both of our candidate models are expected to break down at low momentum transfer because they do not include nuclear effects such as nuclear excitations and collective resonances. In other analyses which fit models to CCQE data, bins which are dominated by low momentum transfer events are excluded~\cite{mb-ccqe-wroclaw}. In this analysis we have not followed any such bin masking procedure. Arguably, to obtain a realistic value of the model parameters, one should only fit the model in its stated region of validity. However, the main focus of this analysis is to obtain central values and errors for the T2K oscillation analysis, where the cross section model is used for all regions of phase-space, so some pragmatism is required.

\section{NuWro as a validation tool for new interaction models}
\label{sec:nuwro}
The NuWro Monte Carlo generator for neutrino interactions has been developed over the past $\sim$10 years at the University of Wroc\l{}aw~\cite{Golan:2012wx}. It was the first MC generator to have an implementation of the Benhar SF~\cite{sf} and the Nieves 2p2h model included~\cite{nieves, nievesExtension}, and served as the benchmark for the \neut development of both models. The implementation of the SF model in NuWro was based on the code written for Reference~\cite{Ankowski:2007uy} and subsequently optimized for NuWro. The Nieves model implementation in NuWro used a series of lookup tables for the 2p2h cross-section as a function of leptonic variables for various nuclear targets and neutrino species so is very similar as in \neut, although it has since been improved to use a more general formalism which depends on a number of nuclear response functions which can be extracted from the Nieves code, and therefore reduces the number of lookup tables required. The same generic model~\cite{multinucleon} was used to simulate the initial and final hadronic states in NuWro as was used in \neut. For both the SF and Nieves 2p2h models, NuWro and \neut are in good agreement, which provides a useful validation of the \neut implementations of these models.

\section{External datasets}
\label{sec:extdata}
Four datasets are used in the CCQE fits presented in this work: the \mb neutrino~\cite{mb-ccqe-2010} (2010) and antineutrino \cite{mb-ccqe-antinu-2013} (2013) results; and the \minerva neutrino~\cite{minerva-nu-ccqe} (2013) and antineutrino~\cite{minerva-antinu-ccqe} (2013) results. All experimental details and information about these results, which is reproduced here, are taken from the references cited above unless otherwise stated.

The single-differential cross section results are given in terms of \qqqe, the four-momentum transfer calculated from lepton kinematics under the quasi-elastic hypothesis, which is calculated using the equations:
\begin{equation} 
  \Enu^{\mathrm{QE}} = \frac{2M'_{n}E_{\mu}-(M'^{2}_{n}+m_{\mu}^{2}-M^{2}_{p})}{2(M'_{n}-E_{\mu}+\sqrt{E_{\mu}^{2}-m_{\mu}^{2}}\cos{\theta_{\mu}})},
  \label{eq:ccqe-enuqe}
\end{equation}
\begin{equation}
  \qqqe = -m_{\mu}^{2}+2\Enu^{\mathrm{QE}}(E_{\mu}-\sqrt{E_{\mu}^{2}-m_{\mu}^{2}}\cos{\theta_{\mu}}),
  \label{eq:ccqe-q2qe}
\end{equation}
where $E_\mu$ is the muon energy; $M_{n}$, $M_{p}$ and $m_{\mu}$ are the neutron, proton and muon masses, respectively; and $M'_{n} = M_{n} - V$ where $V$ is the binding energy of carbon assumed in the analysis\footnote{Note that the binding energy $V$ is just the value assumed when calculating \qqqe, so we must use the same value as the experiments when producing comparable \qqqe distributions, but it need not be consistent with the binding energy used in our simulation.}. For both \mb datasets and the \minerva neutrino dataset, $V = 34$\,MeV; for the \minerva antineutrino dataset, $V = 30$\,MeV.

In the \mb analysis, \qqqe is calculated from the unfolded $T_{\mu}$ and $\cos\theta_{\mu}$ distributions. The \minerva analysis unfolds the \qqqe distribution calculated with the reconstructed $p_{\mu}$ and $\cos\theta_{\mu}$ values. The errors on the \qqqe distributions for both experiments include the uncertainties relating to the muon reconstruction, so should cover the difference in the method used to produce the \qqqe cross section results. We note that the main results of our analysis use the \mb double-differential results only, so there is no possible tension from differences between the methods used to produce \qqqe distributions.

\subsection{\mb neutrino}
The \mb CCQE data has been released as a double-differential cross section as a function of $(T_{\mu}, \cos\theta_{\mu})$, where $T_{\mu}$ is the kinetic energy of the outgoing muon and $\theta_{\mu}$ is the angle between the incoming neutrino and outgoing muon. Differential cross sections were also released as a function of \qqqe or $\Enu^{\mathrm{QE \; RFG}}$, but the double-differential result was preferred as it is contains the most information and has minimal model dependence. The \mb data release included central values for each bin and the diagonal elements of the shape-only covariance matrix; correlations between bins were not released. Additionally, the overall normalization uncertainly was given as 10.7\% for neutrino running.

The \mb CCQE cross sections are released as both CCQE-corrected, and CCQE-like measurements. The CCQE-like sample is obtained by selecting events in which a muon was detected with no pions, but no requirement was made on the proton. The CCQE-corrected measurement is produced by subtracting background events (where the primary interaction was not CCQE) based on the NUANCE~\cite{nuanceMC} generator prediction. The dominant background is CC1$\pi^{+}$, and a dedicated sample was used to tune the NUANCE prediction which was used in the background subtraction. It should be noted that the NUANCE CC1$\pi^{+}$ simulation included $\pi$-less $\Delta$ decay. The published signal purity for the neutrino dataset is 77\%.

CCQE-like results are less model dependent than CCQE-corrected results (as they do not rely on the experiment's own MC correction strategy), but make the analysis dependent on the simulation of the background in the \mb detector, which cannot be tuned to the \mb data in the same way \mb's background model could be. CCQE-corrected results are used in this analysis. A downside of using the CCQE-corrected data is the explicit subtraction of $\pi$-less $\Delta$ decay events in the \mb analysis, which forms part of the Nieves \mnn prediction which we treat as signal in the analysis. Unfortunately, there is no obvious way to account for this effect, so we ignore it for the analysis presented. We note that Nieves {\it et al.} also used the CCQE-corrected dataset to compare to their full models~\cite{nieves_MB_nu_2011, nieves_MB_anu_2013}.

\subsection{\mb antineutrino}
The \mb antineutrino data has been released in the same format as the neutrino mode data. Again, the double-differential CCQE-corrected results are used. The overall normalization uncertainty was given as 13.0\% for antineutrino running. This is likely to be strongly correlated with the normalization uncertainty for the neutrino mode data, as the uncertainly comes mostly from the flux normalization uncertainty. However, as this information was not released, no correlation is assumed in this analysis.

The correction strategy for the antineutrino dataset is more complicated than for the neutrino mode sample because of the relatively high $\nu_{\mu}$ contamination in the $\bar{\nu}_{\mu}$ beam, which is the largest background in the antineutrino CCQE sample (\mb is an unmagnetized detector). There is also a large CC1$\pi^{-}$ background, the analogue of the CC1$\pi^{+}$ contamination in the neutrino dataset. Two properties are used to measure the $\nu_{\mu}$ background~\cite{AguilarArevalo:2011sz}: 8\% of $\nu_{\mu}$-induced CC interactions produce no decay electron due to muon-nucleus capture; and the $\nu_{\mu}$-induced CC1$\pi^{+}$ events can be identified independently of $\bar{\nu}_{\mu}$-induced CC1$\pi^{-}$ as most $\pi^{-}$ mesons are absorbed. Unfortunately, this property makes CC1$\pi^{-}$ a bigger background to the CCQE analysis in antineutrino mode, and means that there is no sample with which to directly tune the CC1$\pi^{-}$ production from the NUANCE resonance model, so the neutrino mode CC1$\pi^{+}$ has to be used (as was done for the neutrino mode sample). Other backgrounds are subtracted using the NUANCE interaction model after some tuning and corrections. As a result of the two large backgrounds in the antineutrino sample, the purity of the CCQE-like sample is 61\%, making the correction larger than for the neutrino mode sample.

\subsection{\minerva}
The CCQE datasets from \minerva are released as CCQE-corrected single-differential flux-averaged cross section as a function of \qqqe, where the flux has been averaged over the region $1.5 \leq \Enu \leq 10$\,GeV. There is an additional requirement that $1.5 \leq \Enu^{\mathrm{QE}} \leq 10$\,GeV, with $\Enu^{\mathrm{QE}}$ as defined in Equation~\ref{eq:ccqe-enuqe}. Covariance matrices and central values have been released to perform fits to both shape-only and absolutely normalized neutrino and antineutrino datasets. In this work, the absolutely normalized distributions have been used in the fit.

The correction strategy for the \minerva data is to fit the relative normalizations of simulated background distributions to the data in terms of the recoil energy, energy deposited outside a vertex region (the recoil region), and then subtract the predicted background from the CCQE-like sample. The published purity for the neutrino dataset ranges from 65\% at low \qqqe to 40\% at high \qqqe (with an overall purity of 49\%). The purity for the antineutrino dataset is given as 77\%. The purity is lower for the neutrino analysis because events with a proton from the initial interaction are more complicated to reconstruct than those with a neutron\footnote{The antineutrino analysis has an additional cut requiring no additional (other than the muon) tracks from the vertex, and allows only one isolated energy shower, whereas the neutrino mode analysis allows two~\cite{minerva-antinu-ccqe, minerva-nu-ccqe}.}.

\begin{figure}[htb]
  \centering
    \includegraphics[width=0.9\columnwidth]{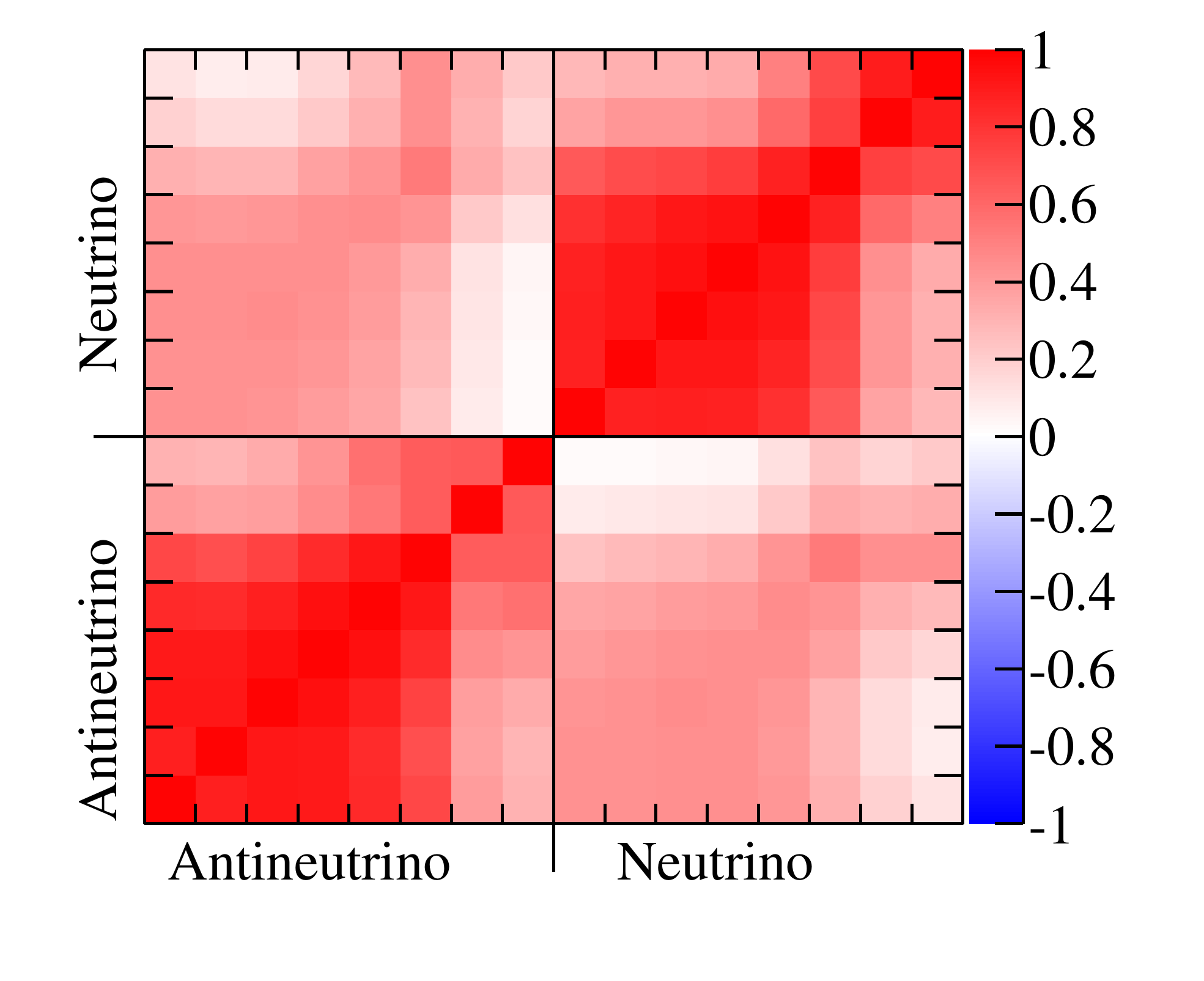}
  \caption{Cross-correlation matrix including both shape and normalization uncertainties for the \minerva neutrino and antineutrino samples. The eight neutrino and eight antineutrino bins shown here correspond to the eight \qqqe bins from the \minerva datasets.}\label{fig:cross-correlation}
\end{figure}

In the \minerva CCQE analyses, the efficiency for selecting events with $\theta_{\mu} > 20^{\circ}$ is very low because the MINOS near detector, downstream of \minerva, is used to tag muons. This introduces a small model dependence on the results because an RFG model was used to correct for the unsampled region of phase-space. The \minerva collaboration subsequently released a distribution where the cross section is measured for CCQE events with $\theta_{\mu} \leq 20^{\circ}$. As this dataset is less model-dependent, it has been used in the fits, and will be consistently used in this analysis. \minerva also made cross-correlations between the neutrino and antineutrino datasets available in a data release after the publication of their CCQE papers. The correlation matrices released include both shape and normalization errors, but it is possible to extract shape-only correlation matrices using the method given in Reference~\cite{teppeiThesis}. The full matrix including both shape and normalization errors included is shown in Figure~\ref{fig:cross-correlation}.

\section{Monte Carlo prediction}
\label{sec:mcpred}
For each of the four experimental results included in the fit, one million CCQE and 2p2h events were generated with \neut for each model using the default parameters given in Tables~\ref{tab:RFG_params} and~\ref{tab:SF_params} and the published flux for each dataset. The flux averaged cross section predictions were produced using the following method:
\begin{enumerate}
  \item For each event apply experiment-specific cuts and, if the event passes, calculate the relevant reconstructed quantity and fill the 1D or 2D event rate histogram.
  \item Calculate the event rate by integrating the MC event rate histogram (flux $\times$ cross section).
  \item Integrate the published flux histogram to get the average flux.
  \item Scale the filled histogram by the event rate divided by the average flux to get the flux averaged cross section per target nucleon.
  \item Divide the content of each bin by the bin width.
\end{enumerate}

\begin{figure}[htb]
  \centering
  \begin{subfigure}[t]{0.9\columnwidth}
    \includegraphics[width=\textwidth]{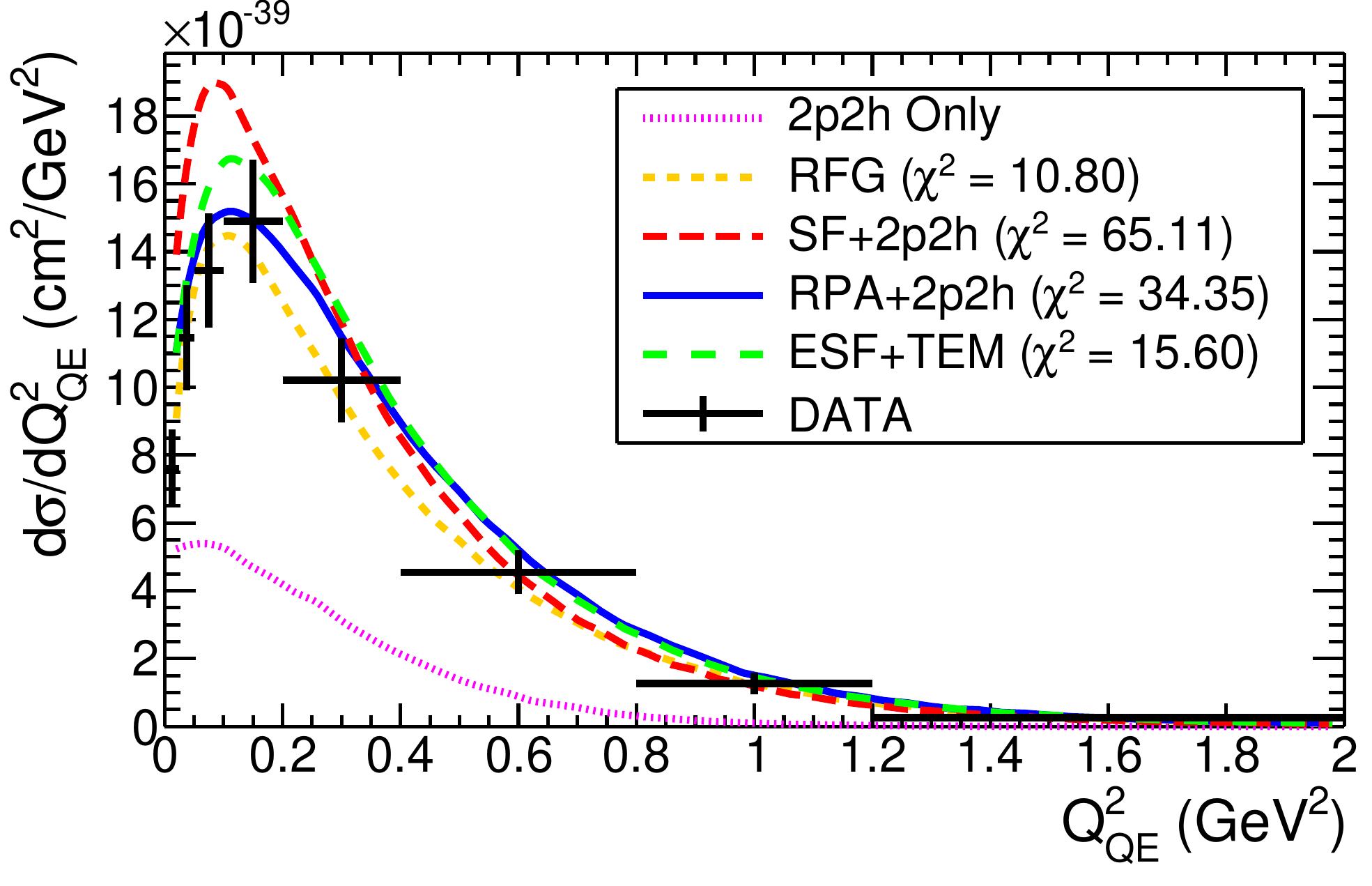}
    \caption{Neutrino}
    \label{subfig:MIN_nu_nominal}
  \end{subfigure}
  \begin{subfigure}[t]{0.9\columnwidth}
    \includegraphics[width=\textwidth]{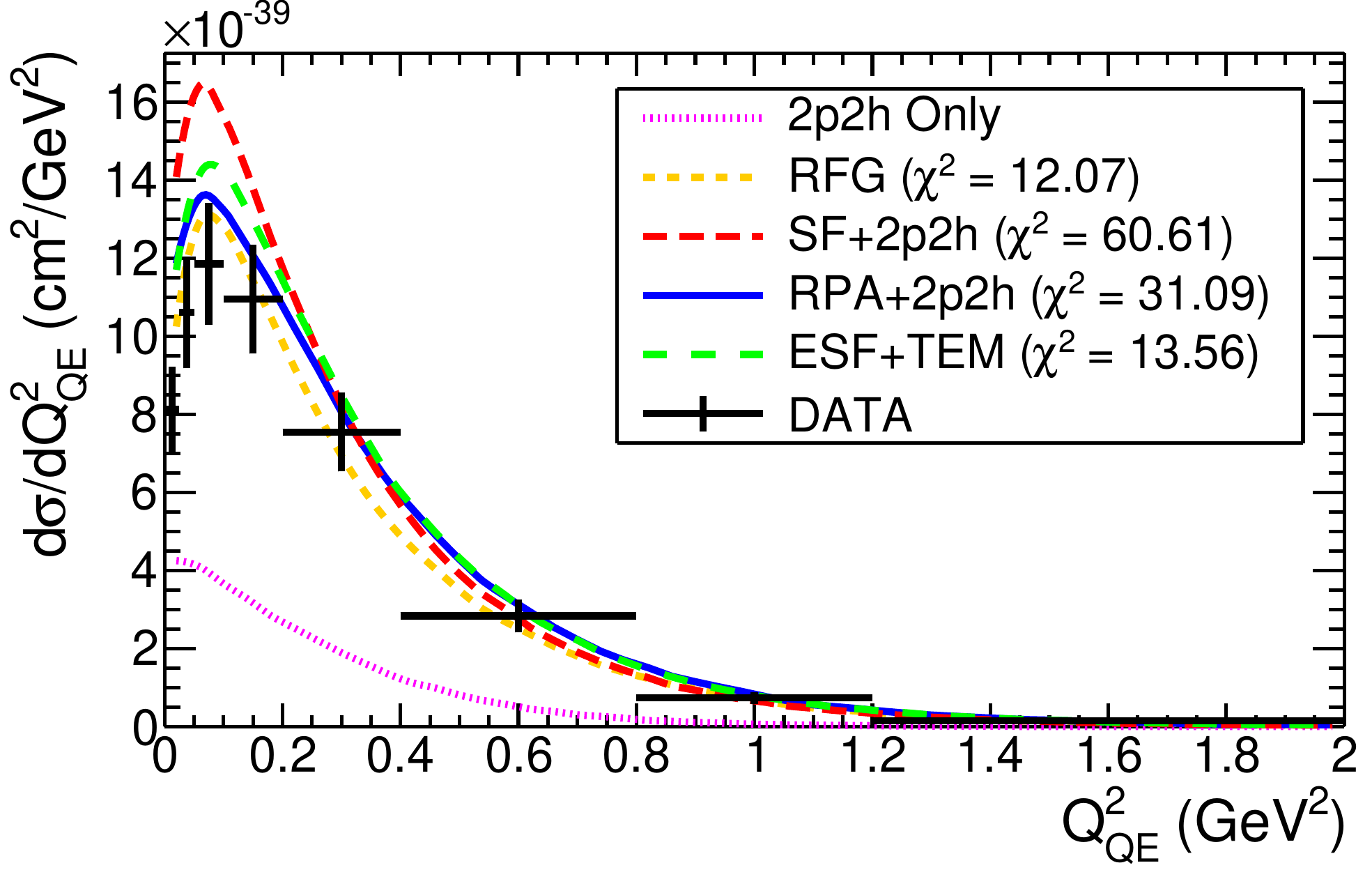}
    \caption{Antineutrino}
    \label{subfig:MIN_antinu_nominal}
  \end{subfigure}
  \caption{Nominal model predictions for the \minerva datasets with \ma$ = 1.01$\,GeV/$c^{2}$ and all other model parameters at their default values.  The relativistic RPA model is shown.}\label{fig:MIN_20deg_nominal}
\end{figure}

\begin{figure}[htb]
  \centering
  \begin{subfigure}[t]{0.9\columnwidth}
    \includegraphics[width=\textwidth]{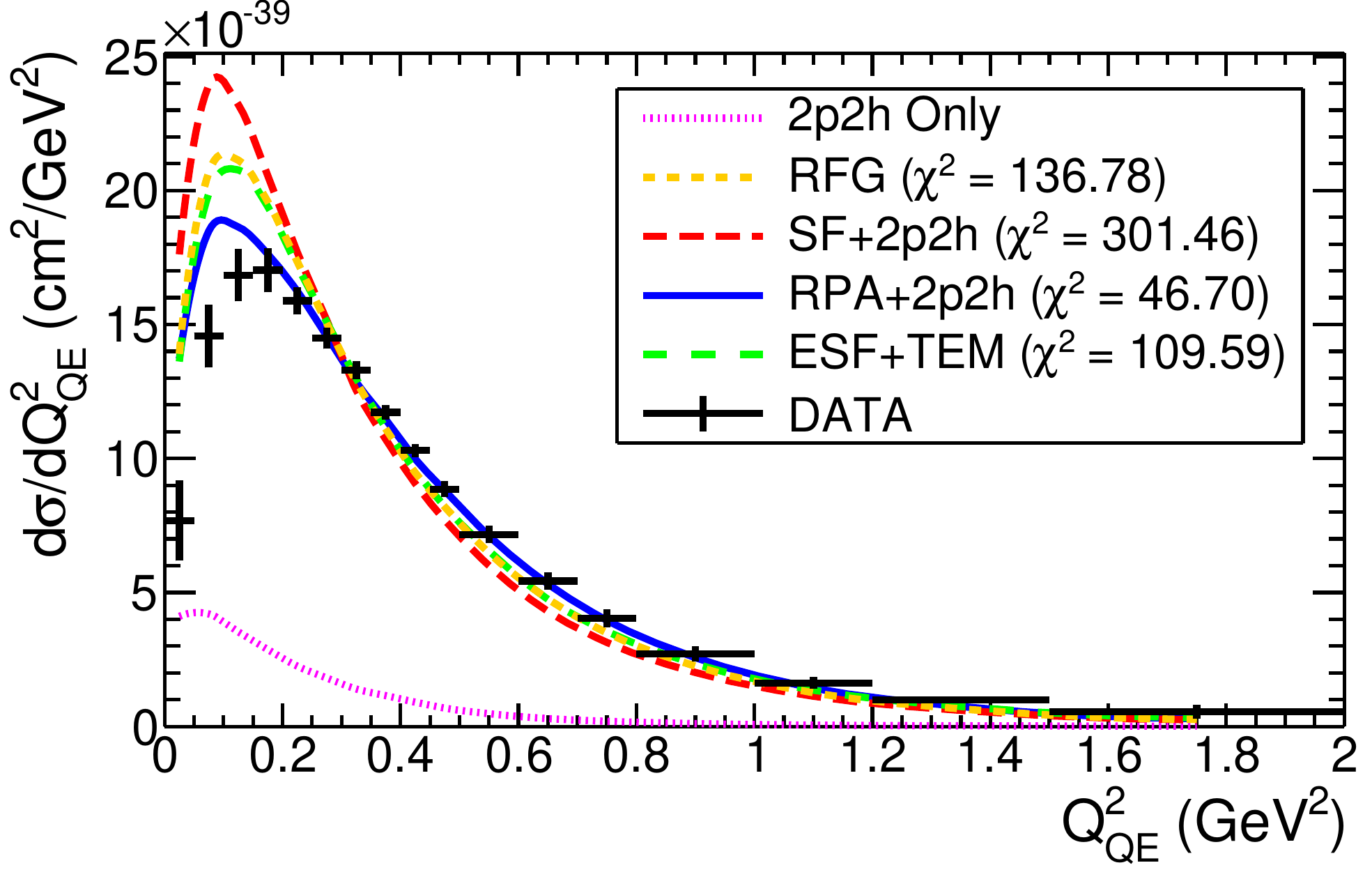}
    \caption{Neutrino}
    \label{subfig:MB_nu_1D_nominal}
  \end{subfigure}
  \begin{subfigure}[t]{0.9\columnwidth}
    \includegraphics[width=\textwidth]{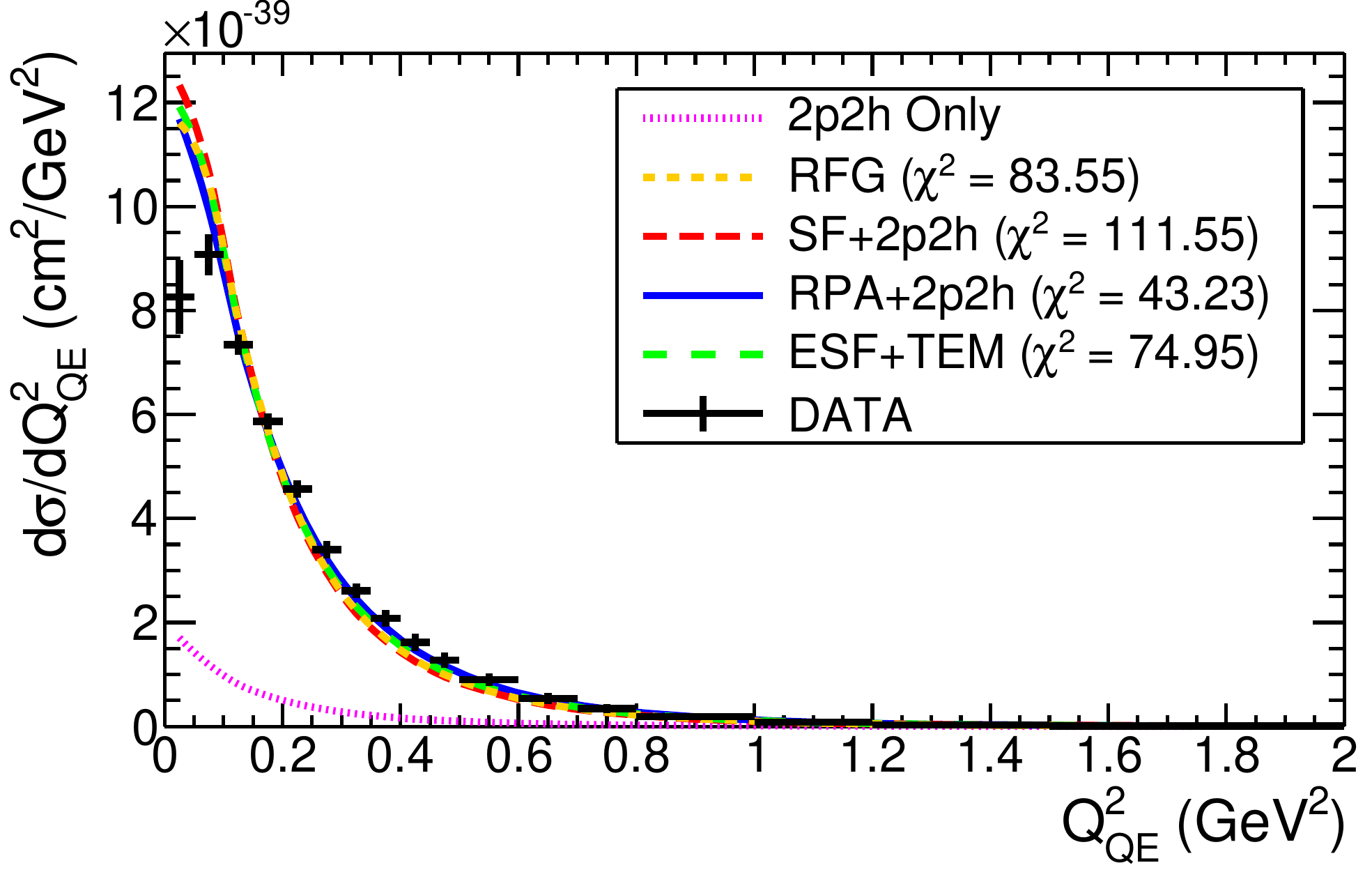}
    \caption{Antineutrino}
    \label{subfig:MB_antinu_1D_nominal}
  \end{subfigure}
  \caption{Nominal model predictions for the \mb single-differential datasets with \ma$ = 1.01$\,GeV/$c^{2}$ and all other model parameters at their default values. The relativistic RPA calculation is shown. Normalization parameters are applied as given in Table~\ref{tab:MB_nominal_fits}.}\label{fig:MB_1D_nominal}
\end{figure}

\begin{figure*}[p]
  \centering
  \begin{subfigure}[t]{0.8\textwidth}
    \includegraphics[width=\textwidth]{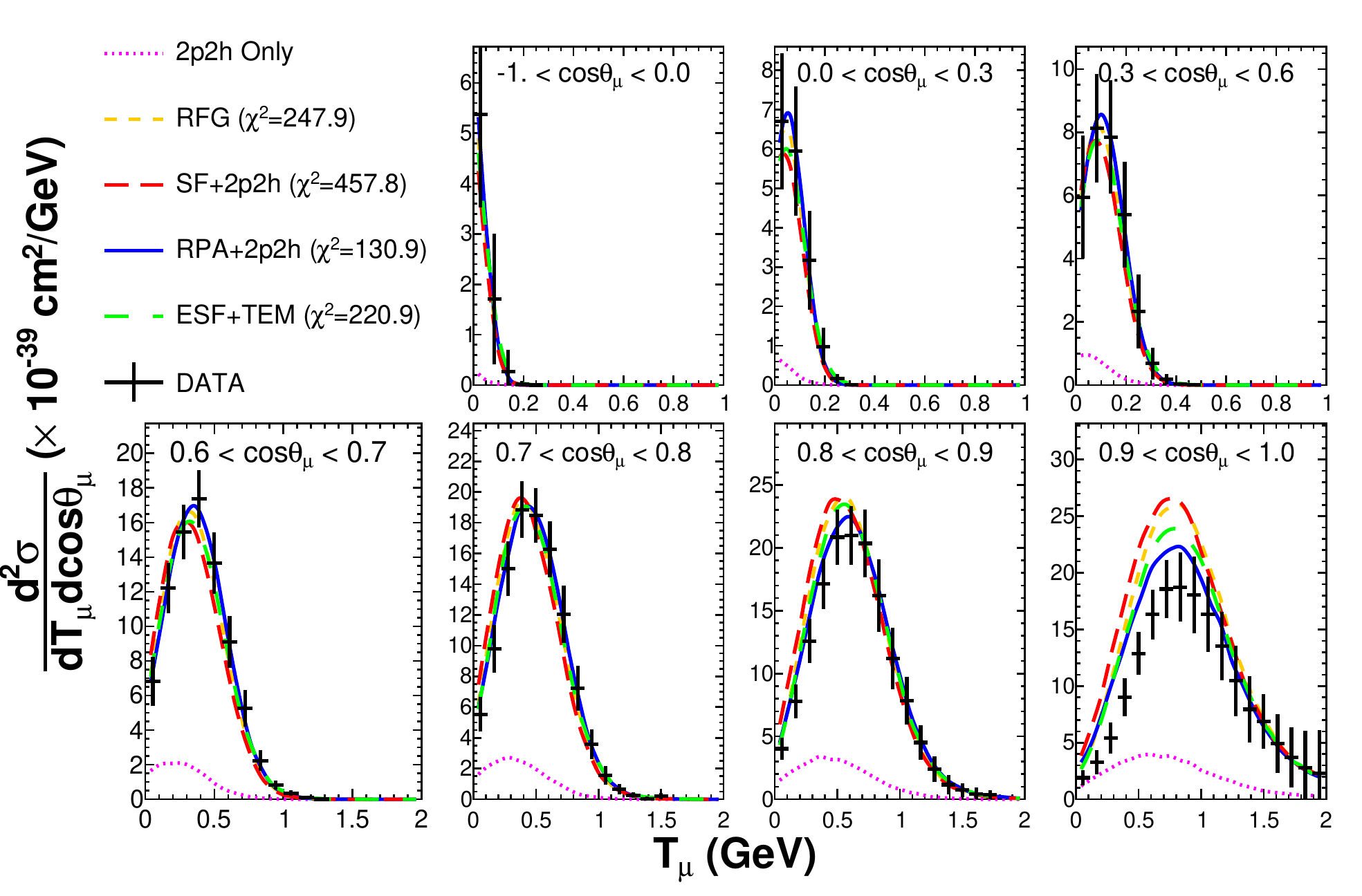}
    \caption{Neutrino}
    \label{subfig:MB_nu_2D_nominal}
  \end{subfigure}
  \begin{subfigure}[t]{0.8\textwidth}
    \includegraphics[width=\textwidth]{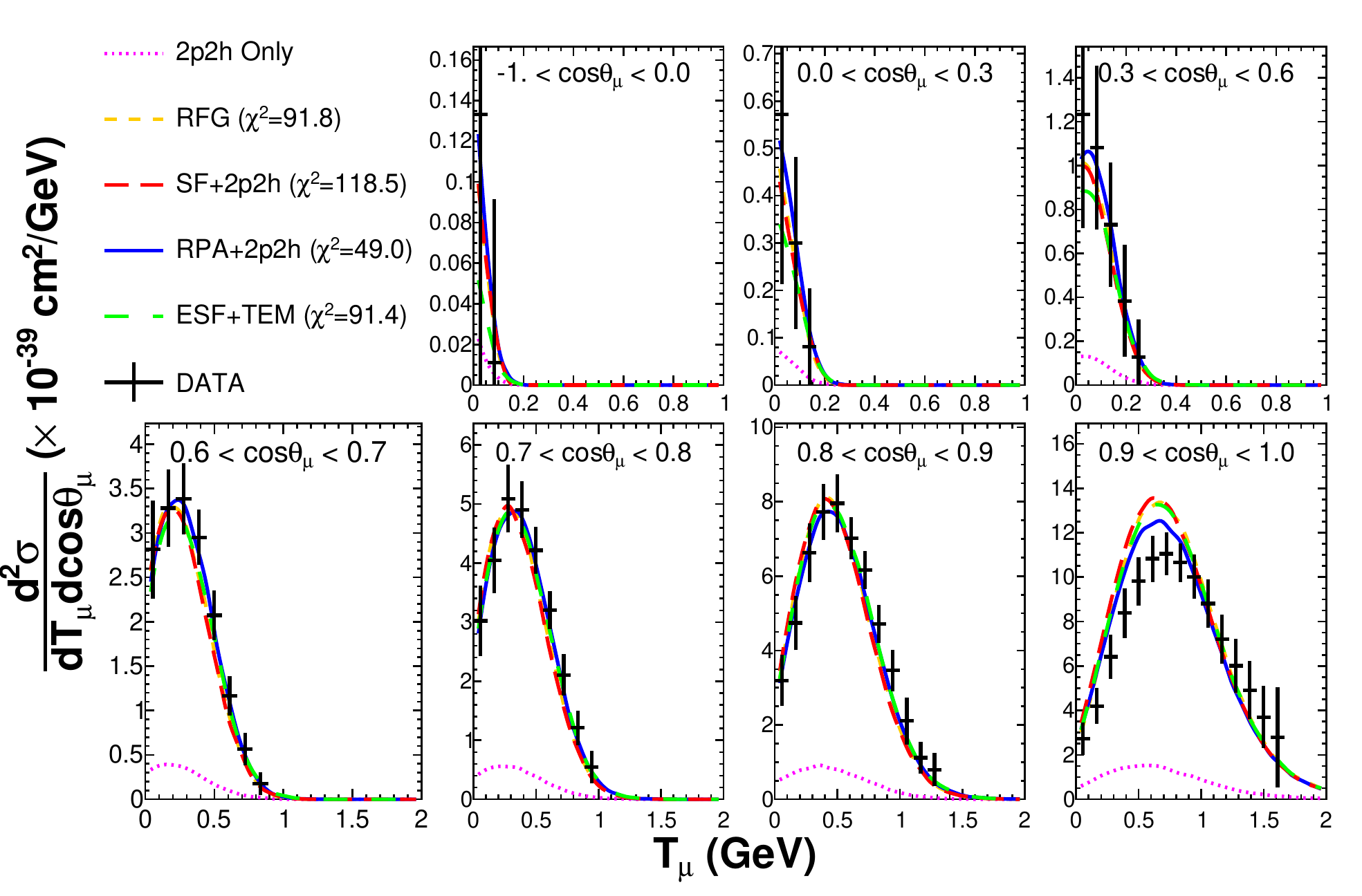}
    \caption{Antineutrino}
    \label{subfig:MB_antinu_2D_nominal}
  \end{subfigure}
  \caption{Nominal model predictions for the \mb double-differential datasets with \ma$ = 1.01$\,GeV/$c^{2}$ and all other model parameters at their default values. The relativistic RPA calculation is shown. Normalization parameters are applied as given in Table~\ref{tab:MB_nominal_fits}.}\label{fig:MB_2D_nominal}
\end{figure*}

The default predictions for a variety of models available in \neut, as well as the data, are shown in Figures~\ref{fig:MIN_20deg_nominal},~\ref{fig:MB_1D_nominal} and~\ref{fig:MB_2D_nominal} for the \minerva, \mb single-differential and \mb double-differential samples, respectively. The Nieves 2p2h contribution is also shown on these plots for reference.

\begin{figure*}[p]
  \centering
  \begin{subfigure}[t]{0.8\textwidth}
    \includegraphics[width=\textwidth]{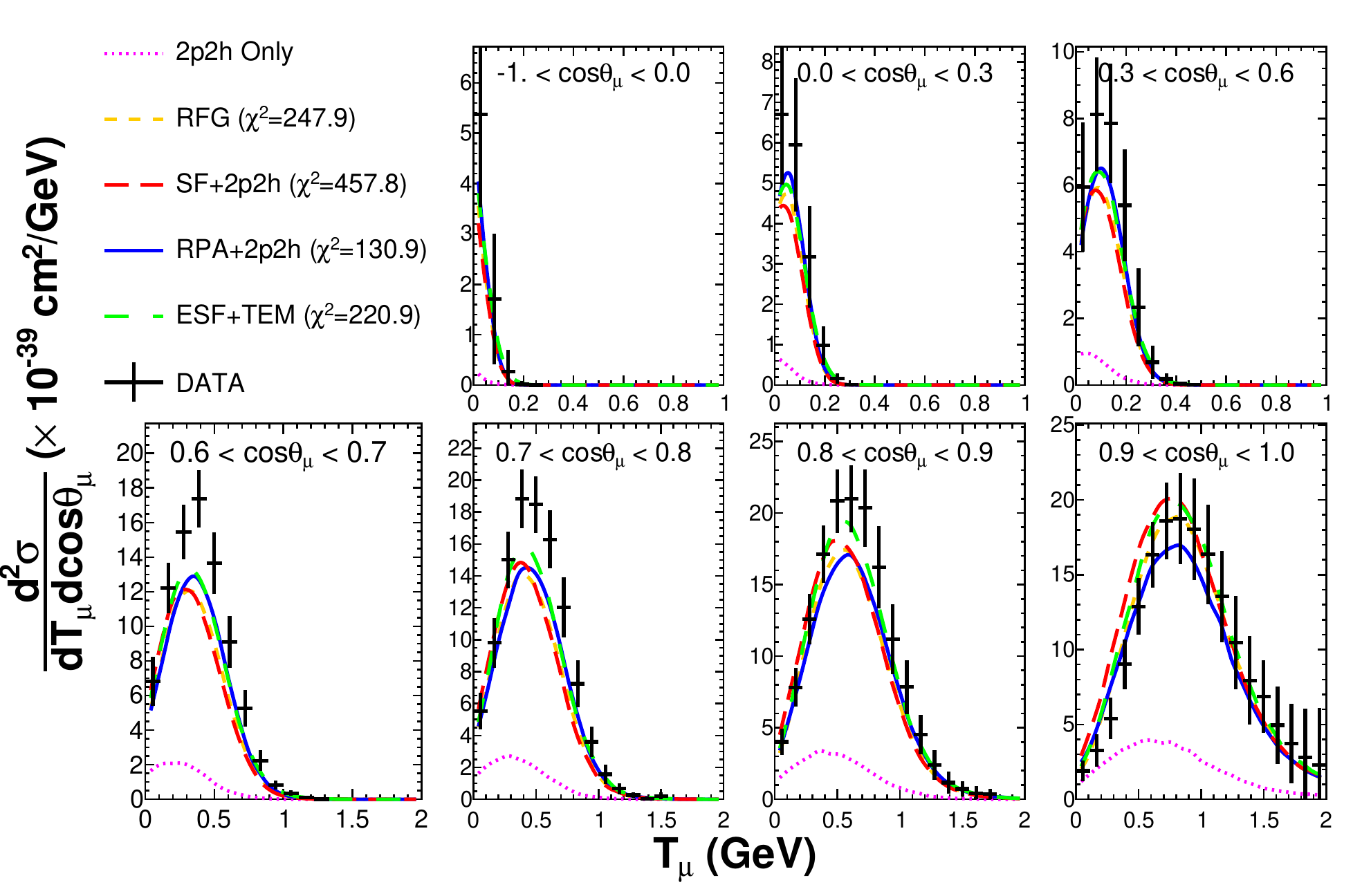}
    \caption{Neutrino}
    \label{subfig:MB_nu_2D_nominal_UNSCALED}
  \end{subfigure}
  \begin{subfigure}[t]{0.8\textwidth}
    \includegraphics[width=\textwidth]{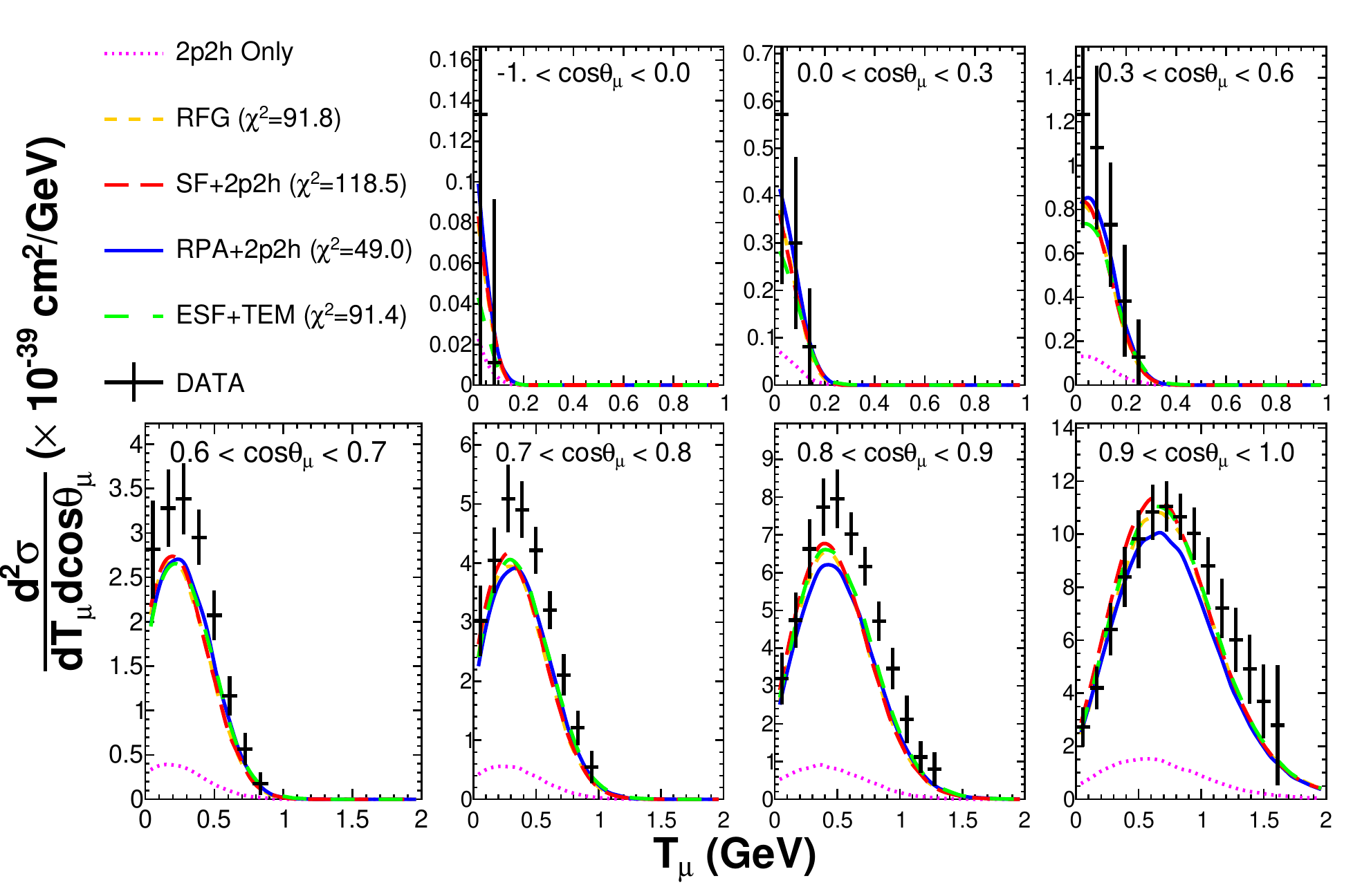}
    \caption{Antineutrino}
    \label{subfig:MB_antinu_2D_nominal_UNSCALED}
  \end{subfigure}
  \caption{Nominal model predictions for the \mb double-differential datasets with \ma$ = 1.01$\,GeV/$c^{2}$ and all other model parameters at their default values. Note that for each model, the relevant \mb normalization parameter has been allowed to vary to minimize the $\chi^{2}$ value, however the scaling factors (given in Table~\ref{tab:MB_nominal_fits}) have not been applied in this figure. The relativistic RPA calculation is shown.}\label{fig:MB_2D_nominal_UNSCALED}
\end{figure*}

To produce a meaningful nominal $\chi^{2}$ for the \mb datasets, it is necessary to fit the \mb normalization parameters. The single and double-differential plots shown in Fig.~\ref{fig:MB_1D_nominal} and~\ref{fig:MB_2D_nominal} are scaled according to the \mb normalization parameter at the best fit point. The best fit values of the pull parameters $\lambda^{\mathrm{MB}}_{\nu}$ and $\lambda^{\mathrm{MB}}_{\bar{\nu}}$ are given in Table~\ref{tab:MB_nominal_fits}. Additionally, the nominal predictions for the \mb double-differential datasets, without the scaling factor applied, are shown in Figure~\ref{fig:MB_2D_nominal_UNSCALED}, which are easier to interpret by eye.


\begin{table}[htb]
 \centering
 {\renewcommand{\arraystretch}{1.2}
\begin{tabular}{c|ccc}
\hline\hline
\multicolumn{2}{c}{Fit type} & $\lambda_{\nu}^{\mathrm{MB}}$ & $\lambda_{\bar{\nu}}^{\mathrm{MB}}$ \\
\hline
\multirow{4}{*}{Neutrino 1D} & RFG & 0.732$\pm$0.007 & --- \\
& SF+2p2h & 0.741$\pm$0.007 & --- \\
& RPA+2p2h & 0.760$\pm$0.007 & --- \\
& ESF+TEM & 0.804$\pm$0.008 & --- \\
\hline
\multirow{4}{*}{Antineutrino 1D} & RFG & --- & 0.805$\pm$0.011 \\
& SF+2p2h & --- & 0.826$\pm$0.011 \\
& RPA+2p2h & --- & 0.774$\pm$0.010 \\
& ESF+TEM & --- & 0.803$\pm$0.011 \\
\hline
\multirow{4}{*}{Neutrino 2D} & RFG & 0.725$\pm$0.011 & --- \\
& SF+2p2h & 0.756$\pm$0.011 & --- \\
& RPA+2p2h & 0.760$\pm$0.011 & --- \\
& ESF+TEM & 0.827$\pm$0.012 & --- \\
\hline
\multirow{4}{*}{Antineutrino 2D} & RFG & --- & 0.808$\pm$0.015 \\
& SF+2p2h & --- & 0.838$\pm$0.015 \\
& RPA+2p2h & --- & 0.802$\pm$0.015 \\
& ESF+TEM & --- & 0.833$\pm$0.015 \\
\hline\hline
\end{tabular}}
 \caption{Table of best fit \mb normalization parameter values for the nominal model comparisons shown in Figures~\ref{fig:MB_1D_nominal} and~\ref{fig:MB_2D_nominal}. The relativistic RPA calculation is shown.}
 \label{tab:MB_nominal_fits}
\end{table}

Note that the double-differential cross section plots shown in Figures~\ref{fig:MB_2D_nominal_UNSCALED} have been rebinned. In the distributions released by \mb, and used in the fits, there are 20 $\cos \theta_{\mu}$ bins uniformly distributed between -1 and 1. For ease of presentation, these have been rebinned and the results are shown in eight $\cos \theta_{\mu}$ slices of varying sizes, where merged bins have been averaged and their errors combined in quadrature.

\FloatBarrier
\begin{widetext}
\section{Fit procedure}
\label{sec:fitprod}

All minimizations are performed using the MIGRAD algorithm of the MINUIT package~\cite{minuit}, using the $\chi^{2}$ statistic:
\begin{align}
  \chi^{2}(\vec{\mathbf{x}}) &= \left\lbrack\sum^{N}_{k=0} \left(\frac{\nu_{k}^{DATA}-\lambda_{\nu}^{-1}\nu_{k}^{MC}(\vec{\mathbf{x}})}{\sigma_{k}} \right)^{2} 
  + \left(\frac{\lambda_{\nu} - 1}{\varepsilon_{\nu}} \right)^{2}\right \rbrack \rightarrow \mathrm{\mb~\nu} \notag\\
  &+ \left\lbrack\sum^{M}_{l=0} \left(\frac{\nu_{l}^{DATA}-\lambda_{\bar{\nu}}^{-1}\nu_{l}^{MC}(\vec{\mathbf{x}})}{\sigma_{l}} \right)^{2} 
  + \left(\frac{\lambda_{\bar{\nu}} - 1}{\varepsilon_{\bar{\nu}}} \right)^{2}\right\rbrack \rightarrow \mathrm{\mb~\bar{\nu}}\notag\\
  &+ \left\lbrack\sum_{i=0}^{16} \sum_{j=0}^{16} \left(\nu_{i}^{DATA} - \nu_{i}^{MC}(\vec{\mathbf{x}})\right)V_{ij}^{-1}\left(\nu_{j}^{DATA} - \nu_{j}^{MC}(\vec{\mathbf{x}})\right)\right\rbrack \rightarrow \mathrm{MINER}\nu\mathrm{A}
\label{eq:chi2_def}
\end{align}
\noindent where $\vec{\mathbf{x}}$ are the model parameters varied in the fit, $\sigma_k$ and $\sigma_l$ are the diagonals of the \mb shape-only covariance matrices for the neutrino and antineutrino results, $V_{ij}$ is the cross-covariance matrix provided by \minerva, and $\lambda_{\alpha}$ and $\lambda_{\beta}$ are the normalization parameters for \mb neutrino and antineutrino datasets, with published normalization uncertainties of $\varepsilon_{\alpha}$ (10.7\%) and $\varepsilon_{\beta}$ (13.0\%)\footnote{Note that the \minerva normalization uncertainty is included in the covariance matrix, so also contributes a penalty term to the fit.}.

Fits to individual datasets only include the relevant terms from the $\chi^{2}$ definition in Equation~\ref{eq:chi2_def}, and fits to single \minerva datasets neglect cross-correlations (the summation is only over the relevant eight bins).

\end{widetext}
\subsection{Parameter Goodness-of-Fit (PGoF) test}\label{sec:pgof}
Standard goodness-of-fit tests, such as the Pearson $\chi^{2}_{\mathrm{min}}$ test used as an example here, test the agreement between prediction and data; however, some issues can arise with their use in global fits, as discussed in Reference~\cite{pgof-2003}. The basic problem is that much of the data will have limited power to constrain any one parameter, but agree well with the prediction regardless of the parameter values. These data will add little to the $\chi^{2}$, but contribute another degree of freedom. Thus the $\chi^{2}_{\mathrm{min}}$ found may be deceptively good despite not agreeing well with those parts of the dataset that actually have power to constrain key parameters. It is also possible that a dataset with a large number of datapoints (such as \mb) which does agree well with a model may hide disagreements with other datasets included in a global fit for which fewer datapoints are available (such as \minerva); again, the key problem is a dilution of the $\chi^{2}$.

This problem is worsened in the case of datasets for which correlations between datapoints have not been included, where the $\chi^{2}/$DOF can be much less than 1, such as is the case for \mb. Looking at the Pearson $\chi^{2}_{\mathrm{min}}$ test statistic is not very illuminating when fitting to both \mb and \minerva datasets. 

The PGoF is a more rigorous test proposed in Reference~\cite{pgof-2003} for fitting to global datasets and has been used extensively in sterile neutrino literature~\cite{conrad12GF, kopp13}, where there are often contradictory results coming from different experiments, and the fitters are fitting to many different experiments which are sensitive to different parameters. It is also referred to as the Likelihood Ratio Test in both statistics and other HEP literature. The PGoF test statistic is given by
\begin{equation}
  \chi_{\mbox{\scriptsize{PGoF}}}^{2}(\vec{\mathbf{x}}) = \chi^{2}_{\mathrm{tot}}(\vec{\mathbf{x}}) - \sum^{D}_{r=1} \chi^{2}_{r, \: \mathrm{min}}(\vec{\mathbf{x}}),
  \label{eq:pgof-test-statistic}
\end{equation}
where $\vec{\mathbf{x}}$ are the parameters floated in the fits, $D$ is the number of datasets, $\chi^{2}_{\mathrm{tot}}$ is the minimum $\chi^{2}$ in a fit to all $D$ subsets of the data, and $\chi^{2}_{r, \: min}$ is the minimum $\chi^2$ obtained in a fit to the $r$th subset of the data. The PGoF test statistic forms a $\chi^{2}$ distribution with the number of degrees of freedom
\begin{equation*}
  P_{\mbox{\scriptsize{PGoF}}} = \sum_{r=1}^{D}P_{r} - P_{\mathrm{tot}},
\end{equation*}
where $P_r$ and $P_{\mathrm{tot}}$ are the number of degrees of freedom for each fit.

The aim of the PGoF is to test the compatibility of the different datasets in the framework of the model. Put simply, it tests whether the best fit parameter values to subsets of the data pulls the fit parameters far from the best fit values found when fitting to all of the data. If different subsets favor very different values, then those subsets are not compatible in the framework of the model (though individually each may be able to find parameter combinations which produce a good fit).

A further advantage of the PGoF test in the situation where some of the data lacks correlations is that the number of degrees of freedom come from the number parameters varied in the fits, not from the number of bins that dataset contributes, which mitigates against the $\chi^{2}_{\mathrm{min}}/\ndof << 1$ issue.

The PGoF test still assumes that the datasets follow a $\chi^{2}$ distribution, but allows for a lower effective number of degrees of freedom. This assumption is not quantitatively correct due to the aforementioned lack of correlations in the \mb CCQE data.  The $p$-values returned should be taken with the caution that they highlight tensions between datasets, but are not to be interpreted in the same manner as they would if all correlations were reported.

\section{Fit results}\label{sec:results}
\subsection{Fake data studies}\label{sec:ccqe_fake_data}
The fitter was validated in two ways. Firstly, Asimov fake datasets~\cite{asimov} were produced to estimate the size of the errors that would be produced from the fit and used as a sanity check of the real fit results. The Asimov tests also provide a very basic test of the fitting framework developed for this analysis. Secondly, pull studies were performed to check that the $\chi^{2}$ definition given in Equation~\ref{eq:chi2_def} is an unbiased estimator of the parameter central values and errors. For all parameters, the biases were less than 10\% across the entire parameter range allowed in the fit, so we conclude that the fitter behaves well.

\subsection{Combined fit}
\label{sec:combined-fit}
\begin{figure}[htb]
  \centering
  \begin{subfigure}[t]{0.9\columnwidth}
    \includegraphics[width=\textwidth]{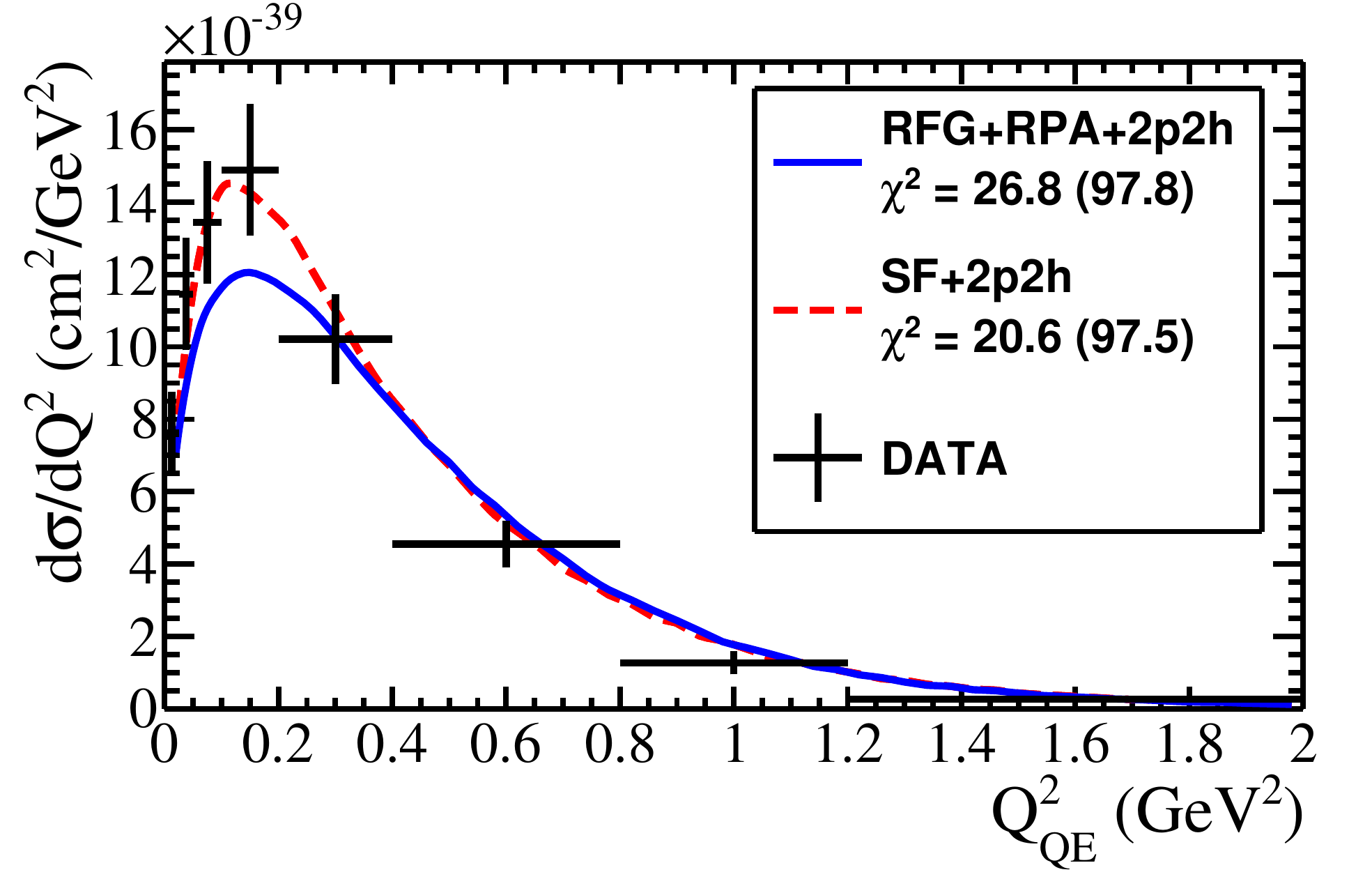}
    \caption{Neutrino}
    \label{subfig:MIN_nu_jointFit}
  \end{subfigure}
  \begin{subfigure}[t]{0.9\columnwidth}
    \includegraphics[width=\textwidth]{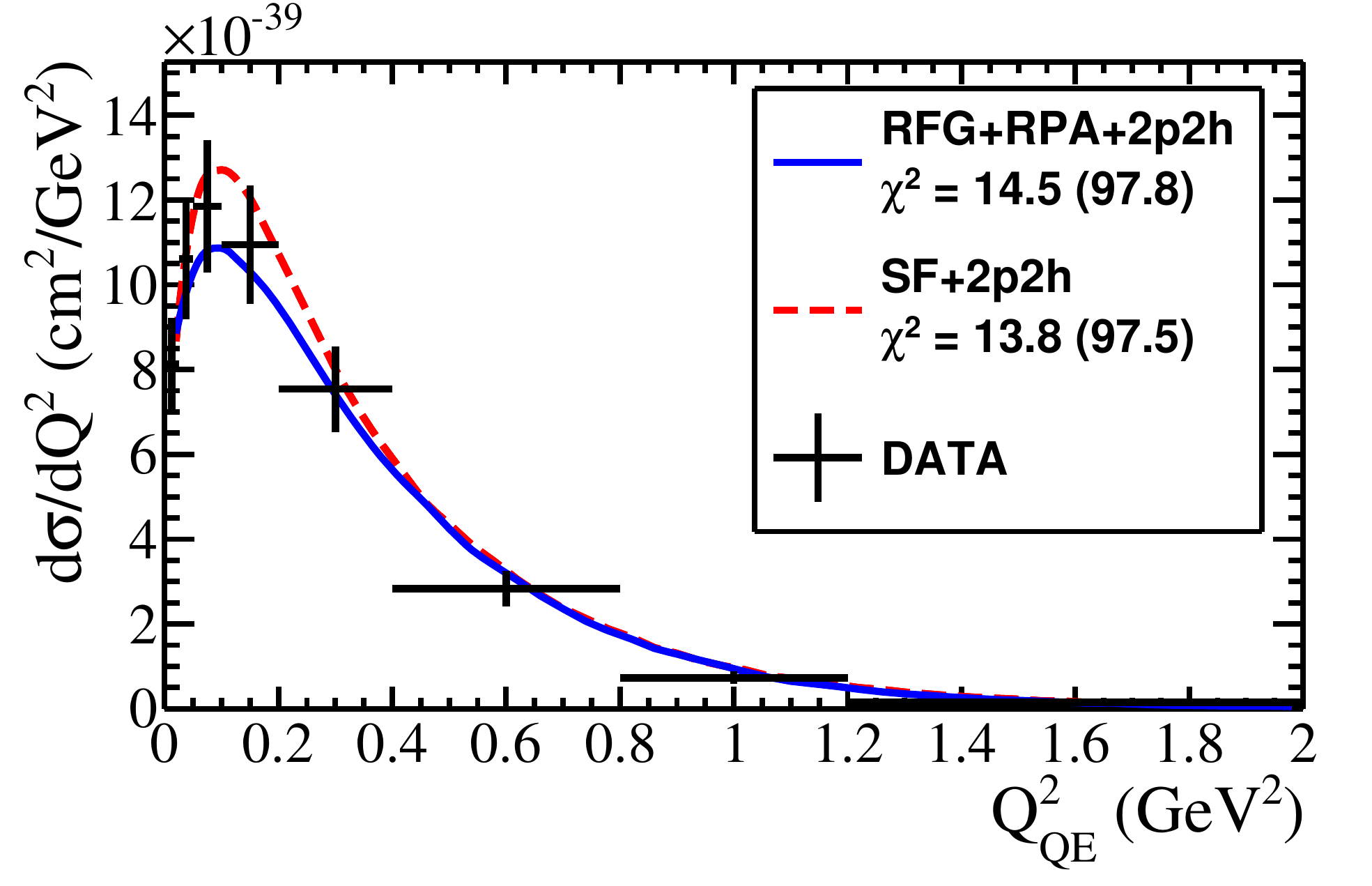}
    \caption{Antineutrino}
    \label{subfig:MIN_antinu_jointFit}
  \end{subfigure}
  \caption{Comparison of the best fit from the combined fits detailed in Table~\ref{tab:combinedSummary} with the \minerva datasets used in the fit. The $\chi^{2}$ values in the legend are the contribution from each dataset at the best fit point (and the total $\chi^{2}_{\mathrm{min}}$ for the combined fit).}\label{fig:MIN_20deg_jointFit}
\end{figure}

\begin{figure*}[p]
  \centering
  \begin{subfigure}[t]{0.8\textwidth}
    \includegraphics[width=\textwidth]{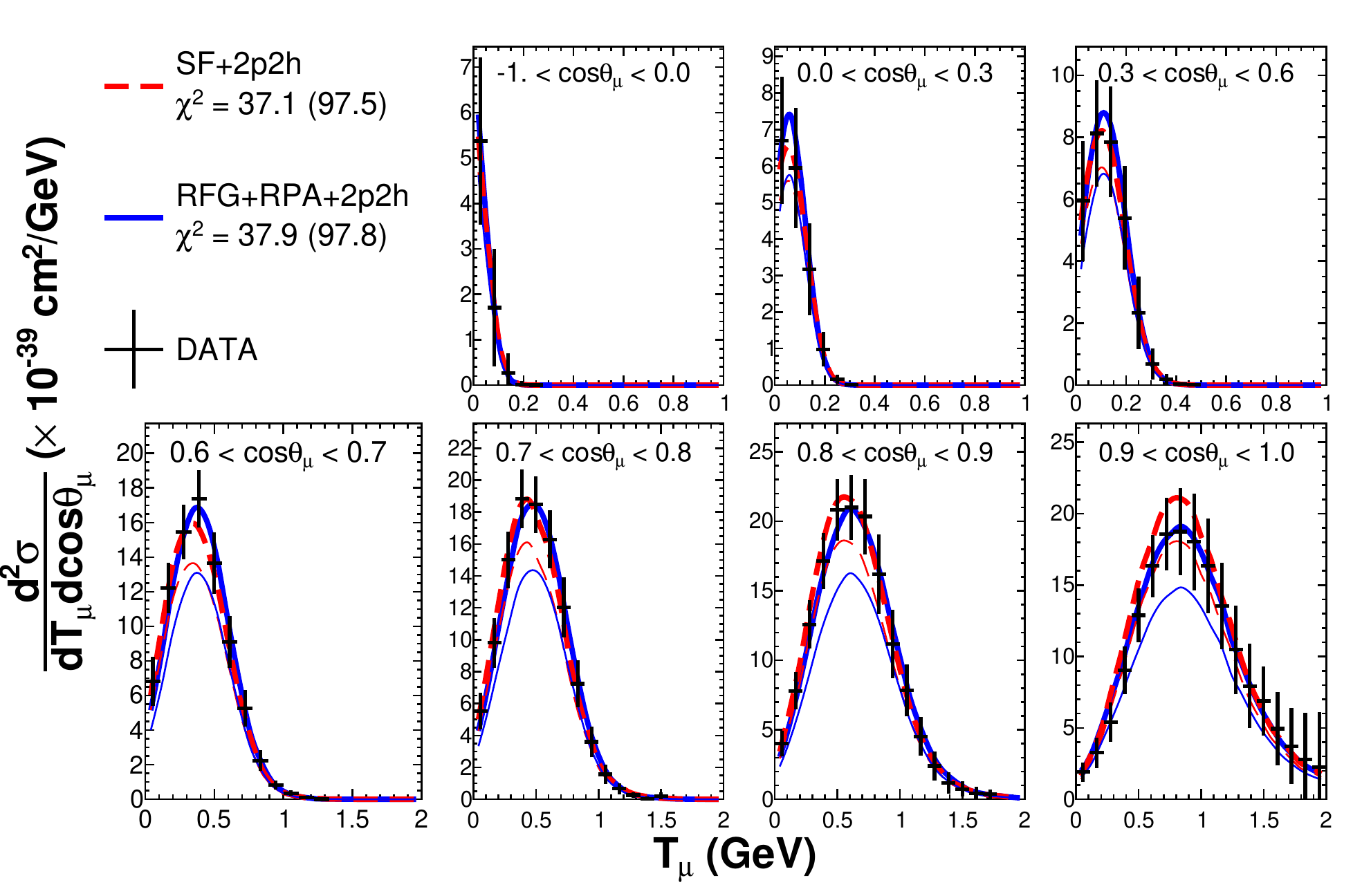}
    \caption{Neutrino}
    \label{subfig:MB_nu_2D_jointFit}
  \end{subfigure}
  \begin{subfigure}[t]{0.8\textwidth}
    \includegraphics[width=\textwidth]{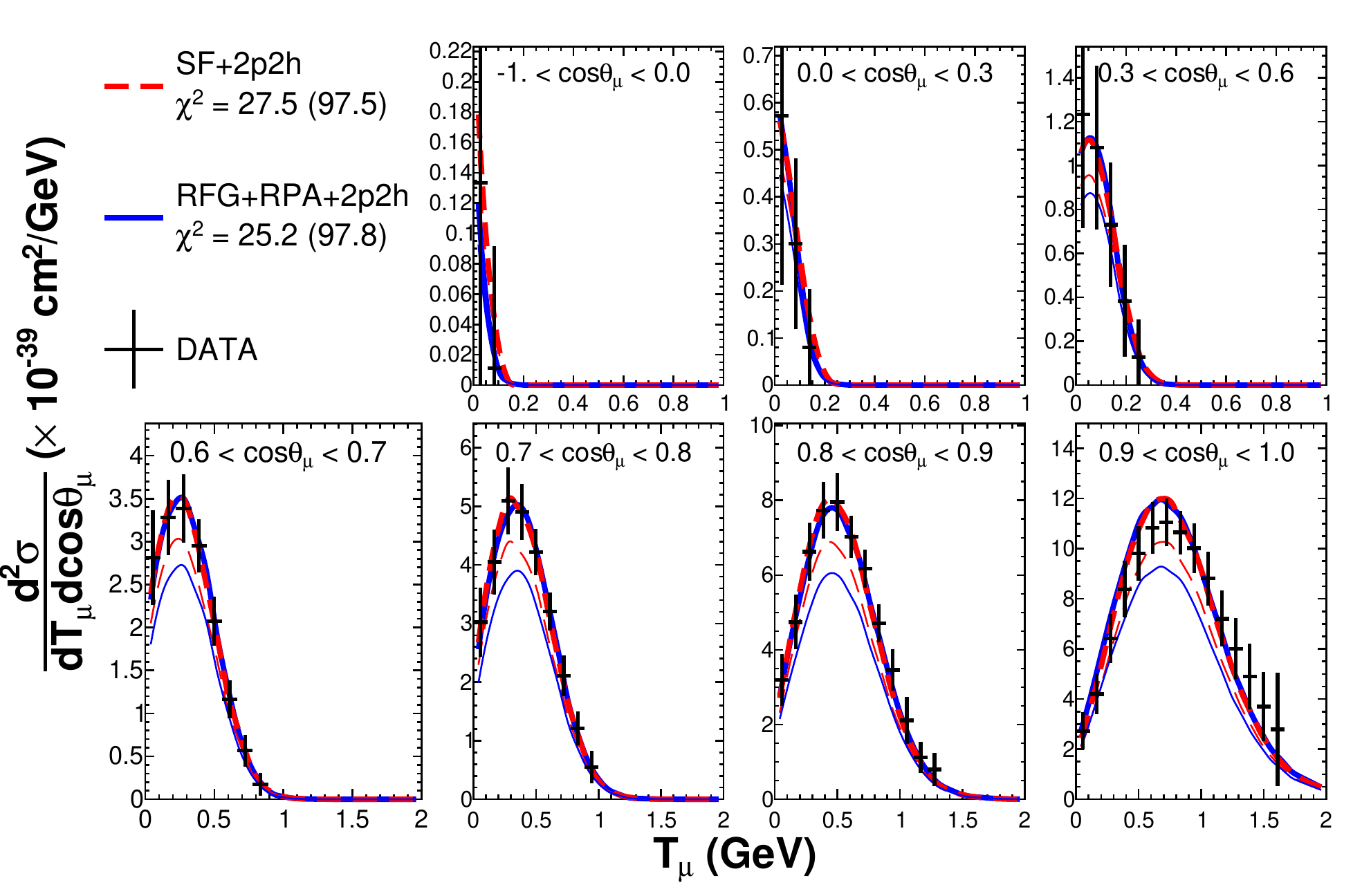}
    \caption{Antineutrino}
    \label{subfig:MB_antinu_2D_jointFit}
  \end{subfigure}
  \caption{Comparison of the best fit from the combined fits detailed in Table~\ref{tab:combinedSummary} with the \mb double-differential datasets used in the fit. The $\chi^{2}$ values in the legend are the contribution from each dataset at the best fit point (and the total $\chi^{2}_{\mathrm{min}}$ for the combined fit). The thick lines have the \mb normalization factors applied (given in Table~\ref{tab:combinedSummary}), while the thin lines do not, to indicate the large pulls on these parameters.}\label{fig:MB_2D_jointFit}
\end{figure*}

The results for the combined fits to all four datasets are given for both relativistic and non-relativistic RFG+RPA+2p2h models and the SF+2p2h model in Table~\ref{tab:combinedSummary}.  The best fit distributions are compared with data for \minerva in Figure~\ref{fig:MIN_20deg_jointFit}, and for \mb in Figure~\ref{fig:MB_2D_jointFit}. Relativistic RPA is used in the figures, as this was the best fit of the two RPA models available. In the legends of these figures, each line is given two $\chi^{2}$ values, the contribution from that dataset to the $\chi^{2}_{\mathrm{min}}$ in the combined fit, and the total $\chi^{2}_{\mathrm{min}}$ in parentheses. Note that in Figure~\ref{fig:MIN_20deg_jointFit}, the contributions from \minerva are calculated for the individual datasets, which necessarily ignores cross-correlations and makes these numbers slightly misleading. Explicitly, $\chi^{2}_{\mathrm{M{\nu}A \: total}} \neq \chi^{2}_{\mathrm{M{\nu}A \: \nu}} + \chi^{2}_{\mathrm{M{\nu}A \: \bar{\nu}}}$ due to cross-correlations, so the values shown in the figure should be treated with caution.
\begin{table*}[htb]
\centering
{\renewcommand{\arraystretch}{1.2}
\begin{tabular}{c|cccccc}
\hline\hline
Fit type & $\chi^{2}$/\ndof & \ma (GeV/$c^{2}$) & 2p2h norm. (\%) & \pf (MeV/$c$) & $\lambda_{\nu}^{\mathrm{MB}}$ & $\lambda_{\bar{\nu}}^{\mathrm{MB}}$ \\
\hline
RFG+rel.RPA+2p2h & 97.8/228 & 1.15$\pm$0.03 & 27$\pm$12 & 223$\pm$5 & 0.79$\pm$0.03 & 0.78$\pm$0.03 \\
RFG+non-rel.RPA+2p2h & 117.9/228 & 1.07$\pm$0.03 & 34$\pm$12 & 225$\pm$5 & 0.80$\pm$0.04 & 0.75$\pm$0.03 \\
SF+2p2h & 97.5/228 & 1.33$\pm$0.02 & 0 (at limit) & 234$\pm$4 & 0.81$\pm$0.02 & 0.86$\pm$0.02 \\
\hline\hline
\end{tabular}}
\caption{Best fit parameter values for the fits to all datasets simultaneously.}\label{tab:combinedSummary}
\end{table*}

It is clear from Figures~\ref{fig:MIN_20deg_jointFit} and~\ref{fig:MB_2D_jointFit} that \mb is not completely dominating the fits, as might be expected given the large number of bins in each of the \mb datasets. Indeed, these fits exploit the fact that, without correlations, $\chi^{2}_{\mathrm{MB}} \approx \chi^{2}_{\mathrm{M{\nu}A}}$. It is also clear that neither model fits all of the datasets perfectly at the best fit point, which is not reflected by the reduced $\chi^{2}$ values of 97.5/228 and 97.8/228 for the SF+2p2h and RFG+rel.RPA+2p2h models, respectively. As the \mb public data release lacks bin-to-bin correlations, the $\chi^{2}_{\mathrm{MB}}$ contributions are not as large as would be expected for the number of bins contributed. This may explain why so many theoretical models are able to find good agreement with the \mb CCQE data.

In all fits, it was observed that the \mb normalization values tended to be suppressed for both neutrino and antineutrino datasets indicating that the MC underestimated the published data by 20--30\% (10--20\%) for the RFG+RPA+2p2h (SF+2p2h) models. It is not possible to accurately determine the favored \minerva normalization as the normalization uncertainty is included in the published covariance matrix, but the output distributions show that the MC normalization is approximately equal to the data normalization.


Because of the large pulls on the \mb normalization parameters, shape-only fits were also performed (see Reference~\cite{my_thesis} for further details). It was found that the best fit parameters were not significantly changed, indicating that there is not a significant bias to the other parameters caused by the large pulls on the \mb normalization parameters. A recent reanalysis of the \minerva flux~\cite{minerva_new_flux} results in an increase to the normalization of previous \minerva cross section results including the CCQE samples used in this analysis. Although these updated datasets are not included in this work, we note that as the results were found to be largely unchanged in a shape-only fit, the main results will not be significantly affected. Additionally, results from fits to individual datasets, and to various combinations of datasets can be found in Reference~\cite{my_thesis}.

\subsection{PGoF results}\label{sec:pgof_results}
Using the PGoF test defined in Section~\ref{sec:pgof}, it is possible to test the compatibility between different subsets of the data. Tables~\ref{tab:PGoF_RFG_non-relativistic},~\ref{tab:PGoF_RFG_relativistic}, and~\ref{tab:PGoF_SF} show a breakdown of the four datasets used in the the combined fits for each initial CCQE model assumption. The Standard Goodness of Fit (SGoF) for each row is determined using Pearson's $\chi^{2}_{\mathrm{min}}$ test, where $\chi^{2}_{\mathrm{min}}$ is found by minimizing the function given in Equation~\ref{eq:chi2_def}, including only the terms for the relevant datasets. The PGoF test is found by subtracting $\chi^{2}_{\mathrm{min}}$ for each of the constituent datasets from the minimum of the combined dataset. The formulae for calculating the PGoF test statistic $\chi^2_{\mathrm{PGoF}}$ are given explicitly in Table~\ref{tab:PGoF_definitions}. The $\chi^{2}_{\mathrm{min}}$ for each dataset is again determined by minimizing the function given in Equation~\ref{eq:chi2_def} with only the relevant terms included.

\begin{table}[htb]
\centering
    {\renewcommand{\arraystretch}{1.2}
        \begin{tabular}{c|c}
        \hline\hline
        &  $\chi^{2}_{\mathrm{PGoF}}$ \\
        \hline
        All & $\chi^{2}_{\mathrm{ALL}} - \chi^{2}_{\mathrm{MB \: \nu}} - \chi^{2}_{\mathrm{MB \: \bar{\nu}}} - \chi^{2}_{[\mathrm{M{\nu}A \: \nu \: + \: \bar{\nu}}]}$ \\
        \minerva & $\chi^{2}_{[\mathrm{M{\nu}A \: \nu \: + \: \bar{\nu}}]} - \chi^{2}_{\mathrm{M{\nu}A \: \nu}} - \chi^{2}_{\mathrm{M{\nu}A \: \bar{\nu}}}$ \\
        \mb & $\chi^{2}_{[\mathrm{MB \: \nu \: + \: \bar{\nu}}]} - \chi^{2}_{\mathrm{MB \: \nu}} - \chi^{2}_{\mathrm{MB \: \bar{\nu}}}$ \\
        $\nu$ & $\chi^{2}_{[\mathrm{MB \: \nu \: + \: M{\nu}A \: \nu}]} - \chi^{2}_{\mathrm{MB \: \nu}} - \chi^{2}_{\mathrm{M{\nu}A \: \nu}}$ \\
        $\bar{\nu}$ & $\chi^{2}_{[\mathrm{MB \: \bar{\nu} \: + \: M{\nu}A \: \bar{\nu}}]} - \chi^{2}_{\mathrm{MB \: \bar{\nu}}} - \chi^{2}_{\mathrm{M{\nu}A \: \bar{\nu}}}$ \\
        M$\nu$A vs MB & $\chi^{2}_{\mathrm{ALL}} - \chi^{2}_{[\mathrm{MB \: \nu \: + \: MB \: \bar{\nu}}]} - \chi^{2}_{[\mathrm{M{\nu}A \: \nu \: + \: \bar{\nu}}]}$ \\
        $\nu$ vs $\bar{\nu}$ & $\chi^{2}_{\mathrm{ALL}} - \chi^{2}_{[\mathrm{MB \: \nu \: + \: M{\nu}A \: \nu}]} - \chi^{2}_{[\mathrm{M{\nu}A \: \bar{\nu} \: + \: MB \: \bar{\nu}}]}$ \\
        \hline\hline
     \end{tabular}}
     \caption{Explicit formulae for calculating the $\chi^{2}_{\mathrm{PGoF}}$ test statistics for each of the subsets of the data investigated. Each $\chi^2$ value listed in this table denotes the $\chi^2$ at the minimum.}\label{tab:PGoF_definitions}
\end{table}

In each fit, the \ma, 2p2h normalization, \pf, and any \mb normalization terms are allowed to float.
\begin{table}[htb]
\centering
   {\renewcommand{\arraystretch}{1.2}
   \begin{tabular}{c|cccc}
     \hline\hline          
     & $\chi^{2}_{\mathrm{min}}$/\ndof & SGoF (\%) &  $\chi^{2}_{\mathrm{PGoF}}$/\ndof & PGoF (\%) \\
     \hline
     All & 117.9/228 & 100.00 & 25.3/6 & 0.03 \\
     \minerva & 30.3/13 & 0.42 & 0.4/3 & 93.09 \\
     \mb & 65.7/212 & 100.00 & 3.4/3 & 33.09 \\
     $\nu$ & 69.1/142 & 100.00 & 12.7/3 & 0.53 \\
     $\bar{\nu}$ & 46.1/83 & 99.97 & 10.4/3 & 1.55 \\
     M$\nu$A vs MB & 117.9/228 & 100.00 & 21.9/3 & 0.01 \\
     $\nu$ vs $\bar{\nu}$ & 117.9/228 & 100.00 & 2.6/3 & 45.12 \\
     \hline\hline
   \end{tabular}}
\caption{PGoF results for various subsets of the data for the RFG+non-rel.RPA+2p2h model.}\label{tab:PGoF_RFG_non-relativistic}
\end{table}

One subtlety must be kept in mind when analysing the results in Tables~\ref{tab:PGoF_RFG_non-relativistic},~\ref{tab:PGoF_RFG_relativistic}, and~\ref{tab:PGoF_SF}: the PGoF test is only appropriate for statistically independent datasets. This makes the interpretation difficult for \minerva, where cross-correlations are provided and used in the fits. Whenever a subset of data includes both \minerva $\nu$ and $\bar{\nu}$ datasets, the fits include cross-correlations, but if only one dataset is included, they are not. This means that two of the rows in each table give slightly unreliable results: ``\minerva'', and ``$\nu$ vs $\bar{\nu}$''. In each case, the $\chi^{2}$ function for the combined dataset includes cross-correlations, and the $\chi^{2}$ functions for the subdivided dataset does not. The issue is most obvious in Table~\ref{tab:PGoF_RFG_relativistic}, where the ``$\nu$ vs $\bar{\nu}$'' row gives a negative PGoF $\chi^{2}$. These values are still useful as a comparison between models and to give a rough idea of compatibility between datasets, but the exact values must be treated with caution.

\begin{table}[htb]
\centering
   {\renewcommand{\arraystretch}{1.2}
   \begin{tabular}{c|cccc}
     \hline\hline
     & $\chi^{2}_{\mathrm{min}}$/\ndof & SGoF (\%) &  $\chi^{2}_{\mathrm{PGoF}}$/\ndof & PGoF (\%) \\
     \hline
     All & 97.8/228 & 100.00 & 17.9/6 & 0.66 \\
     \minerva & 23.4/13 & 3.74 & 1.0/3 & 79.03 \\
     \mb & 58.6/212 & 100.00 & 2.0/3 & 57.69 \\
     $\nu$ & 62.6/142 & 100.00 & 16.1/3 & 0.11 \\
     $\bar{\nu}$ & 38.5/83 & 100.00 & 6.1/3 & 10.75 \\
     M$\nu$A vs MB & 97.8/228 & 100.00 & 15.9/3 & 0.12 \\
     $\nu$ vs $\bar{\nu}$ & 97.8/228 & 100.00 & -3.3/3 & 100.00 \\
     \hline\hline
   \end{tabular}}
\caption{PGoF results for various subsets of the data for the RFG+rel.RPA+2p2h model.}\label{tab:PGoF_RFG_relativistic}
\end{table}

\begin{table}[htb]
\centering
  {\renewcommand{\arraystretch}{1.2}
  \begin{tabular}{c|cccc}
    \hline\hline
    & $\chi^{2}_{\mathrm{min}}$/\ndof & SGoF (\%) &  $\chi^{2}_{\mathrm{PGoF}}$/\ndof & PGoF (\%) \\
    \hline
    All & 97.5/228 & 100.00 & 41.1/6 & 0.00 \\
    \minerva & 12.6/13 & 47.75 & 1.0/3 & 79.49 \\
    \mb & 50.2/212 & 100.00 & 6.5/3 & 8.92 \\
    $\nu$ & 54.8/142 & 100.00 & 25.1/3 & 0.00 \\
    $\bar{\nu}$ & 34.1/83 & 100.00 & 8.5/3 & 3.61 \\
    M$\nu$A vs MB & 97.5/228 & 100.00 & 34.6/3 & 0.00 \\
    $\nu$ vs $\bar{\nu}$ & 97.5/228 & 100.00 & 8.5/3 & 3.59 \\
    \hline\hline
\end{tabular}}
\caption{PGoF results for various subsets of the data for the SF+2p2h model.}\label{tab:PGoF_SF}
\end{table}

The PGoF test highlights the incompatibility of the various datasets within the framework of the SF+2p2h and both RFG+RPA+2p2h models, despite the apparent goodness of fit when only considering $\chi^{2}_{\mathrm{min}}/\ndof$. The level of agreement given in the final column of Tables~\ref{tab:PGoF_RFG_non-relativistic},~\ref{tab:PGoF_RFG_relativistic}, and~\ref{tab:PGoF_SF} should be interpreted as the level of agreement between the datasets included in that row. For example, it is clear that for all models considered the agreement found between the \minerva and \mb datasets (which include both neutrino and antineutrino samples) have the lowest level of agreement as shown by the ``M$\nu$A vs MB'' row. In contrast, the level of agreement between the neutrino and antineutrino datasets (which include the \minerva and \mb samples) show relatively good agreement, indicating that fitting to the neutrino and antineutrino datasets separately produces similar best fit parameter values.

It is clear from Table~\ref{tab:PGoF_SF} that the SF+2p2h model does not fit the various datasets well, the poor PGoF statistics indicate that the datasets favor very different parameter values when fit separately. This is particularly true for any fits involving the \mb neutrino dataset, though there is no {\it a priori} reason to exclude this dataset and improve the fit results. The PGoF tests for RFG+RPA+2p2h using both relativistic and non-relativistic RPA, shown in Tables~\ref{tab:PGoF_RFG_non-relativistic} and~\ref{tab:PGoF_RFG_relativistic}, show much better compatibility between experiments than SF+2p2h. There is still a considerable amount of tension, which is largely due to differences between \minerva and \mb. Because of the relatively poor consistency between datasets for the SF+2p2h model compared with RFG+rel.RPA+2p2h, the latter model is a better choice as the default model for T2K oscillation analyses.

\subsection{Rescaling parameter errors}\label{sec:scale-errors}
Assuming Gaussian statistics, 1$\sigma$ errors on a single fit parameter are defined by the parameter value for which $\chi^{2} = \chi^{2}_{\mathrm{min}} + 1$~\cite{pdg2014}. MINUIT uses this assumption when calculating the errors at the minimum, which were included with the best fit values for the combined fit in Table~\ref{tab:combinedSummary}. However, as well as motivating the use of the PGoF test, the lack of bin correlations from \mb also means that Gaussian statistics no longer work as expected when estimating parameter errors.

There is a large body of literature looking at how this problem affects fits to parton density distributions, where global fits include a large number of datasets, many of which did not provide bin correlations~\cite{pumplin2000_pdfs, stump_2001_pdfs, collins_2001_pdfs}. A summary of the work of one PDF fitting group is given in Reference~\cite{pumplin2000_pdfs} and was used as a guide here. Their solution for producing reasonable parameter error estimates is to inflate the value of the $\Delta \chi^{2}$ used to define the 1$\sigma$ parameter errors, although no generic solution is offered for defining that value. In the case of the PDF fits in Reference~\cite{pumplin2000_pdfs}, the $\Delta \chi^{2}$ used was very large, $\sim$100, although it should be kept in mind that many more datasets are used in that fit than in the current work.

The PGoF gives a value for the incompatibility between the datasets: how much the $\chi^{2}$ increases between the best fit points of each experiment and the best fit point for the combined dataset. The PGoF value can therefore be used as a measure of how much the errors have to be inflated to cover the difference between the best fit parameter values from the combined fit and the best fit values of individual datasets, this is shown explicitly in Equation~\ref{eq:pgof-rescaling}, where the value used to define the 1$\sigma$ error is given by $\Delta \chi^{2}$, and the rescaling parameter is given by $r$.
\begin{align}
  \Delta \chi^{2} &= \chi^{2}_{\mbox{\scriptsize{PGoF}}}/\ndof \notag \\
  r &= \sqrt{\chi^{2}_{\mbox{\scriptsize{PGoF}}}/\ndof}
  \label{eq:pgof-rescaling}
\end{align}
Note that this PGoF rescaling procedure does not modify the correlations between parameters, it simply rescales the error on each parameter.

\begin{table}[htb]
\centering        
 {\renewcommand{\arraystretch}{1.2}
  \begin{tabular}{c|cccc}
    \hline\hline
    Fit type & $\chi^{2}$/\ndof & \ma (GeV/$c^2$) & 2p2h (\%) & \pf (MeV/$c$) \\
    \hline
    Unscaled & \multirow{2}{*}{97.8/228} & 1.15$\pm$0.03 & 27$\pm$12 & 223$\pm$5~ \\
    PGoF scaling & & 1.15$\pm$0.06 & 27$\pm$27 & 223$\pm$11 \\
    \hline\hline
\end{tabular}}
\caption{The final errors for the RFG+rel.RPA+2p2h parameters. Note that the scaled errors should be used by any analyses which use these results.}\label{tab:final-errors}
\end{table}

There is some ambiguity over which PGoF statistic to use, the `All' or `M$\nu$A vs MB' row of Table~\ref{tab:PGoF_RFG_relativistic}, with $\chi^{2}_{\mbox{\scriptsize{PGoF}}}/\ndof$ values of 17.9/6 and 15.9/3 respectively. The more conservative value is from the `M$\nu$A vs MB' (because the greatest differences are between experiments, not between neutrino and antineutrino running), so this is used\footnote{It should also be noted that this rescaling procedure more than covers the difference between neutrino and antineutrino datasets.}. To be explicit, we multiply the parameter errors from MINUIT by $r = \sqrt{15.9/3}\approx2.3$ based on this statistic, as shown in Table~\ref{tab:final-errors}. It can be seen from Table~\ref{tab:final-errors} that the 2p2h normalization is strongly suppressed and, even with the rescaled error, is nearly 3$\sigma$ away from the Nieves nominal model prediction. It is also clear that although the axial mass value preferred in the fit is not as strongly inflated as in fits to MiniBooNE data alone~\cite{mb-ccqe-wroclaw, mb-ccqe-2010, mb-ccqe-antinu-2013}, it is still significantly higher than the value of $\ma \simeq 1$ GeV$/c^{2}$ found by fitting to light target data and pion electroproduction data~\cite{bbba05}, and the inflated 1$\sigma$ error does not cover this difference.

\section{Discussion of the fit results}
\label{sec:discussion}
The results from the fits presented in Section~\ref{sec:combined-fit} show that none of the models which are currently available in \neut describe all of the \ccqe data adequately. In particular, there is a significant difference between \mb and \minerva data which forces the model parameters to compromise between the two, as well as a large change in the normalization for the \mb datasets. Although the \ma value obtained from the fit to the RFG+rel.RPA+2p2h model is lower than that obtained from past fits of the RFG model to \mb data alone (see Reference~\cite{Abe:2015awa} as an example), it is still inconsistent with that obtained in global fits to light target bubble chamber data or high energy heavy target data~\cite{bbba05}.  Additionally, the data requires a large supression of the nominal 2p2h model. The SF+2p2h model, in which the 2p2h component is completely suppressed, requires an inflated \ma value to fit the data. This is unsurprising as the SF model alone does not significantly change the total cross section. Including an RPA calculation appropriate for the SF model is likely to reduce the tension with the 2p2h model, and is likely to change this conclusion significantly, this work will be revisited when such a calculation is available.  Both fits also initially imply that there may need to be additional interactions used that may mimic CCQE interactions or change the shape of the distributions through additional, but currently unmodelled, effects in the nucleus.

There seems to be a shape problem with the 2p2h model, which leads to the suppression of the 2p2h cross section when 2p2h normalization is allowed to vary in the fit. Recall that at the best fit point 2p2h is suppressed to 27\% of the Nieves nominal value as shown in Table~\ref{tab:final-errors}. This suppression is driven by \minerva, which would completely suppress the 2p2h component of the model if \mb were not included in the fit~\cite{my_thesis}. The SF+2p2h model shows the same disagreement, and 2p2h is completely removed at the best fit point. The inability to change the shape of the 2p2h prediction is clearly a deficiency in the current \neut implementation of the Nieves 2p2h model, and much better agreement might be found if more parameters could be varied in the model. A further significant issue is that the CCQE-corrected cross section results from both \minerva and \mb have part of the 2p2h signal region removed as a background ($\pi$-less $\Delta$ decay). It is not clear how this issue should be treated in the fits, and has simply been ignored here, as indeed has been done by the 2p2h model builders~\cite{nieves, martini}. Future cross section measurements should be encouraged to focus on exclusive final states (CC$0\pi$) rather than initial state processes (CCQE), which will avoid such an issue in the future. As previously remarked, the RFG+RPA+2p2h model implemented in \neut is not equivalent to the full Nieves model because the 1p1h component in \neut uses a global, rather than local, Fermi gas nuclear model. Using a global Fermi gas model results in greater interaction strength in the low \qq region than with a local Fermi gas, which is also where the 2p2h model contributes most interaction strength. It is possible that the 2p2h shape issue is due to a conflict with the 1p1h model, and that a more consistent LFG+RPA+2p2h model would resolve this issue.

Although both the RFG+rel.RPA+2p2h model and SF+2p2h model give reasonable agreement with data at the best fit point, it is difficult to trust standard goodness of fit tests as the lack of \mb correlations means that Gaussian statistics no longer work correctly. An alternative measure of the goodness of fit, the PGoF, was used to try to improve the situation. Although the PGoF procedure still assumes Gaussian statistics, it highlights disagreements within the combined dataset by dividing the dataset into subsamples. These disagreements are completely hidden by the standard goodness of fit tests because the \mb $\chi^{2}$ contribution is so low relative to the number of degrees of freedom it contributes to the fit. The PGoF showed that there was considerably better agreement between the best fit parameter values obtained in fits to subsamples of the data for the RFG+rel.RPA+2p2h fit, which gives some confidence to the fit result. For the SF+2p2h model, the fits to subsamples of the data pulled to drastically different parameter values at the best fit points, which is highly undesirable behaviour if the fit results are to be used as prior uncertainties in oscillation analyses, and indicates that the model is a bad fit to the global dataset. But the SF+2p2h model can fit individual datasets well (as is clear in Table~\ref{tab:PGoF_SF}), so should not be discounted completely. 

The lack of reported \mb correlations and non-Gaussian behaviour of the test statistic also means that standard parameter error estimation does not work, and returns smaller parameter errors than are reasonable given the level of disagreement between the datasets used in the fit. An unrealistically tight constraint on cross section parameters would lead to biases in the near detector fit for T2K. To circumvent this problem a PGoF error inflation procedure was defined to ensure that the 1$\sigma$ parameter errors cover the disagreement between the \minerva and \mb datasets. This is a conservative approach, but as no model seems able to describe all of the available data, such an approach was necessary. Such {\it ad hoc} procedures are necessary when incomplete information is available from some of the datasets included in the fit. The lack of information about bin-to-bin correlations for the \mb datasets significantly complicates this analysis, and may significantly change the results. We note that in the literature, many statements about how well various models agree with the \mb datasets are made which assume Gaussian statistics. It is clear that an appreciation of this issue is important for future model comparisons, and that the availability of complete information for new cross section results will be critical for building consistent CCQE models.

\begin{figure}[htb]
  \centering
  \includegraphics[width=0.9\columnwidth]{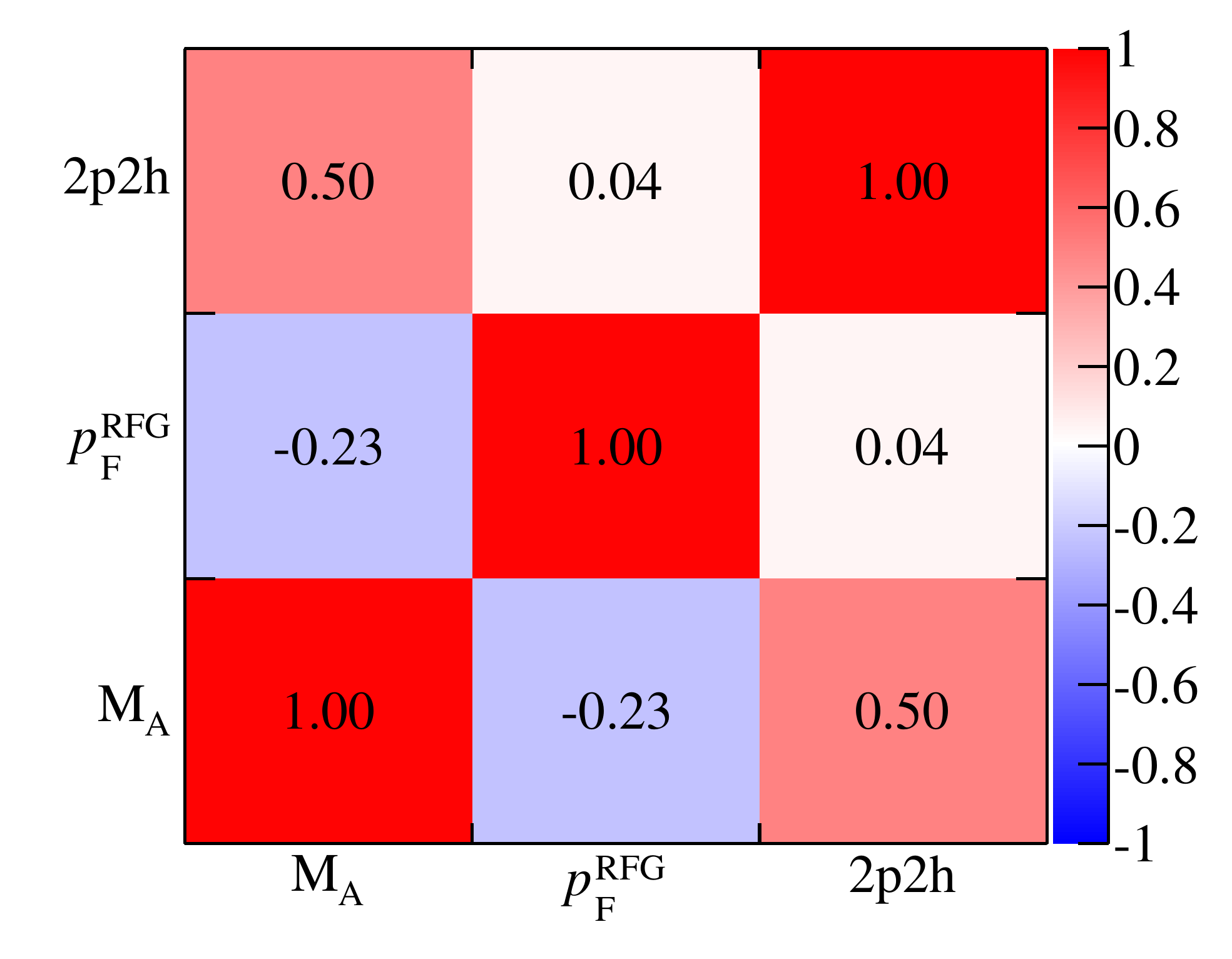}
  \caption{Correlation matrix for the best fit RFG+rel.RPA+2p2h model parameters.}\label{fig:final-correlation}
\end{figure}

For T2K, the results of this fit are part of a larger set of cross section model systematic uncertainties recommended by the NIWG that can be used as prior inputs for the oscillation analyses and various cross section analyses. In this case, the model used for these analyses is the RFG+rel.RPA+2p2h model, since the SF+2p2h model is disfavored in the fits and relativistic RPA is preferred over non-relativistic RPA.  The best fit parameters and uncertainties of the model are given in the second row of Table~\ref{tab:final-errors}, and are correlated according to the matrix shown in Figure~\ref{fig:final-correlation}.

\section{Summary}
\label{sec:summary}
In this paper, we have shown how T2K's NIWG uses previously published CCQE datasets from the \mb and \minerva experiments to test CCQE+2p2h models in the \neut neutrino interaction generator.  For each model, the parameters that describe the data are fit, with both the SGoF and PGoF used to select the model that best describes the data.  In this case, the RFG+rel.RPA+2p2h model is considered the best candidate, with $\ma = 1.15 \pm 0.03$\,GeV/$c^2$, the normalization on the 2p2h model $27\pm12$\%, and $\pf=223\pm5$\,MeV/$c$.  Tensions between the two experiments require an error scaling procedure outlined by the PGoF test, with the final result providing prior inputs into various future T2K analyses.  This is the first time a comprehensive analysis has been performed with these models using a neutrino interaction generator and published, and the first time that such models will be used in an oscillation analysis with full detector simulations~\cite{t2k_numubar_disapp_2015}.

Moving away from the RFG model for \ccqe interactions is an ambitious step for a neutrino experiment as it is a departure from the standard which has been used for decades~\cite{review}. The new models on the market are not perfect, and their implementation into \neut and other neutrino interaction generators will always have technical foibles. However, further theoretical development of these models requires the engagement of the experimental community, and so using them in our simulations is essential to move the field onwards. It is also clear that the current approach of inflating \ma is inadequate, and something better must be done in order to make precision measurements of neutrino oscillation parameters.

The fitting framework developed by the NIWG for this analysis is extensible and the general method for producing cross section errors developed in this work will be used with new \ccqe models and datasets in future, and with new cross section channels entirely, to continue to contrain systematic errors for T2K oscillation and cross section analyses. The results from the \ccqe fits presented here will also help inform the future model development required to fit the data. It is clear that alternative 2p2h models and fundamental parameters in the 2p2h model should be investigated, to see whether the disagreement with the 2p2h shape is telling us something meaningful about the Nieves model. It is also probable that the current RPA model is too inflexible, and this is partially responsible for the disagreement between \mb and \minerva data. Both of these problems may relate to the fact that for several years, the only data available for theorists to use for building models against was from the \mb neutrino dataset, which is difficult to use due to the lack of correlations and the explicit subtraction of $\pi$-less $\Delta$ decay from the CCQE-corrected result. Converging on a new \ccqe model which adequately describes all current and future data is likely to require several iterations between experimentalists and model builders. Confronting all the available models with a variety of data, as has been done in this analysis, and including these models in full Monte Carlo simulations, as will be done in T2K with the output from this analysis, is an important step in this cycle from the experimental side.

\begin{acknowledgments}
The authors would like to thank the members of the T2K Collaboration and the authors of the NuWro generator for their help and support in this analysis. We thank the \minerva and \mb collaborations for assistance in understanding their results and studies beyond the published work. We are grateful to J. Nieves and his collaborators for providing the code required to implement their 2p2h and RPA models in \neut. We are grateful to T. Katori and P. Litchfield for helpful comments on the manuscript. We thank the J-PARC staff for superb accelerator performance and the 
CERN NA61/SHINE collaboration for providing valuable particle production data.
We acknowledge the support of MEXT, Japan; 
NSERC, NRC and CFI, Canada; 
National Science Centre (NCN), Poland; 
MINECO and ERDF funds, Spain; 
SNSF and SER, Switzerland; 
STFC, UK; 
and DOE, USA. 
We also thank CERN for the UA1/NOMAD magnet, 
DESY for the HERA-B magnet mover system, 
NII for SINET4, 
the WestGrid and SciNet consortia in Compute Canada, 
GridPP, UK, 
and the Emerald High Performance Computing facility 
in the Centre for Innovation, UK. 
\end{acknowledgments}

\bibliographystyle{apsrev4-1}
\bibliography{niwgpaperdraft}
\end{document}